\documentclass[preprint,notoc]{JHEP3}
\usepackage{epsfig}

\def\to{\rightarrow}  
  
\def\bi{\begin{itemize}}  
\def\ei{\end{itemize}}  
\def\te{\tilde e}  
  
\def\tu{\tilde u}  
  
\def\tb{\tilde b}  
  
\def\td{\tilde d}

\def\tst{\tilde t}  
\def\ttau{\tilde \tau}  
  
\def\tg{\tilde g}  
\def\tnu{\tilde\nu}  
\def\tell{\tilde\ell}  
\def\tq{\tilde q}  
\def\tw{\widetilde W}  
\def\tz{\widetilde Z}  
\def\alt{\lesssim}  
\def\agt{\stackrel{>}{\sim}}  
\def\be{\begin{equation}}  
\def\ee{\end{equation}}  
\newcommand{\lesim}{\stackrel{\scriptscriptstyle<}{\scriptscriptstyle\sim}}

\title{Yukawa Coupling Unification \\ in Supersymmetric Models}  
  
\author{Daniel Auto, Howard Baer, Csaba Bal\'azs, Alexander Belyaev\footnote  
{On leave of absence from Nuclear Physics Institute, Moscow State University.}   
\\ Department of Physics, Florida State University\\   
Tallahassee, FL, USA 32306\\  
E-mail: \email{auto@hep.fsu.edu},   
\email{baer@hep.fsu.edu}, \email{balazs@hep.fsu.edu},  
        \email{belyaev@hep.fsu.edu}}  
\author{Javier Ferrandis, Xerxes Tata \\  
Department of Physics and Astronomy, University of Hawaii, \\  
Honolulu, HI 96822, USA    \\  
E-mail: \email{javier@phys.hawaii.edu}, \email{tata@phys.hawaii.edu}}  
  
\preprint{\vbox{\hbox{FSU-HEP-030211} \vspace{0.2cm}  
                \hbox{UH-511-1016-03}}}   
  
\abstract{We present an updated assessment of the viability of  
$t-b-\tau$ Yukawa coupling unification in supersymmetric models.  
For the superpotential Higgs mass parameter $\mu >0$, we find  
unification to less than 1\% is possible, but only for  
GUT scale scalar mass parameter $m_{16}\sim 8-20$~TeV,   
and small values of gaugino mass $m_{1/2}\lesim 400$ GeV.   
Such models require that  
a GUT scale mass splitting exists amongst Higgs scalars with  
$m_{H_u}^2<m_{H_d}^2$.    
Viable solutions lead to  
a radiatively generated inverted scalar mass hierarchy, with   
third generation and Higgs scalars being lighter than other sfermions.  
These models have very heavy sfermions, so that unwanted flavor changing  
and $CP$ violating SUSY processes are suppressed, but may suffer from some  
fine-tuning requirements.  
While the generated spectra satisfy $b\to s\gamma$ and $(g-2)_\mu$  
constraints, there exists tension with the dark matter relic density  
unless $m_{16}\lesim 3$ TeV. These models offer prospects  
for a SUSY discovery at the Fermilab Tevatron collider via the search  
for $\tw_1\tz_2\to 3\ell $ events, or via gluino pair production.  
If $\mu <0$,   
Yukawa coupling unification to less than 5\% can occur for   
$m_{16}$ {\it and} $m_{1/2} \agt 1-2$~TeV.  
Consistency of negative $\mu$ Yukawa unified models with $b\to s\gamma$,  
$(g-2)_\mu$, and relic density $\Omega h^2$ all imply very large  
values of $m_{1/2}$ typically greater than about 2.5 TeV, in which case  
direct detection of sparticles may be a challenge even at the LHC.  
}  
  
  
\begin{document}  
  
\section{Introduction}  
\label{sec:intro}  
  
The successful unification of gauge couplings in the  
Minimal Supersymmetric Standard Model (MSSM) at the scale  
$M_{GUT}\simeq 2\times 10^{16}$ GeV provides a compelling hint for the  
existence of some form of a supersymmetric grand unified   
theory (SUSY GUT). SUSY GUT models based on the gauge group $SO(10)$  
are particularly intriguing\cite{review}.   
In addition to unifying gauge couplings,   
\begin{itemize}  
\item they unify all matter of a single generation into the  
16 dimensional spinorial multiplet of $SO(10)$.  
\item The {\bf 16} of $SO(10)$ contains in addition to all SM  
matter fields of a single generation a gauge singlet   
right handed neutrino state which naturally leads to a mass  
for neutrinos. The well-known see-saw mechanism\cite{seesaw}   
implies that if $m_{\nu_\tau}  
\sim 0.03$ eV, as suggested by atmospheric neutrino data\cite{superk},   
then the mass scale associated with $\nu_R$ is very close  
to the $GUT$ scale: {\it i.e.} $M_N\sim 10^{15}$ GeV.  
\item $SO(10)$ explains the apparently fortuitous   
cancellation of triangle anomalies within the SM.  
\item The structure of the neutrino sector of $SO(10)$ models lends  
itself to a successful theory of baryogenesis via   
intermediate scale leptogenesis\cite{leptogen}.  
\end{itemize}  
  
In the simplest $SO(10)$ SUSY GUT models, the SM Higgs doublets are both  
present in a single 10-dimensional Higgs multiplet of $SO(10)$. In these  
models, there exists the additional prediction of Yukawa coupling  
unification for the third generation: $f_t=f_b=f_\tau$, where the  
superpotential of such models contains the term \be \hat{f}\ni  
f\hat{\psi}({\bf 16})^T\hat{\psi} ({\bf 16})\hat{\phi} ({\bf 10})  
+\cdots .  \ee Here, the dots represent possible additional terms  
including higher dimensional Higgs representations which would be  
responsible for the breaking of $SO(10)$ in four dimensional models.  
Realistic models of SUSY $SO(10)$ grand unification in four spacetime  
dimensions are challenging to construct in that one is faced with {\it  
i.}) obtaining the appropriate pattern of $SO(10)$ gauge symmetry  
breaking, leaving the MSSM as the low energy effective field theory,  
{\it ii.}) obtaining an appropriate mass spectrum of SM matter  
fields\cite{fmasses,blazek}, 
and {\it iii.})  avoiding too rapid proton decay,  
which can be mediated by color triplet Higgsinos (the so-called  
doublet-triplet splitting problem)\cite{dtsplit}, though there are  
proposals\cite{dtsplitsol} that lead to large doublet-triplet splitting.  
In addition, the large dimensional Higgs representations needed for GUT  
symmetry breaking appear unwieldy and unnatural, and can even lead to  
models with non-perturbative behavior above the GUT scale.  
  
Recently, progress has been made in constructing SUSY GUT models  
where the grand unified symmetry is formulated in 5 or more spacetime   
dimensions. In such models, the GUT symmetry can be   
broken by compactification of the extra dimension(s) on an appropriate  
topological manifold, such as an $S_1/(Z_2\times Z_2')$ orbifold.   
A variety of models have been proposed for both  
gauge groups $SU(5)$\cite{exdimguts} and $SO(10)$\cite{so10exd}.   
A common feature of such models is that  
symmetry breaking via compactification can alleviate many of the  
problems associated with four dimensional GUTS, while maintaining  
its desirable features such as  
gauge coupling unification, matter   
unification and Yukawa coupling unification.  
In addition, extra dimensional SUSY GUT models can explain  
features of the SM fermion mass spectrum based on  
``matter geography'', {\it i.e.} on whether matter fields exist  
predominantly on one or another of the 3-branes, or in the bulk.  
  
In this paper, we do not adopt any specific SUSY $SO(10)$ GUT model,   
but instead assume that some SUSY GUT model exists, be it  
extra dimensional or conventional, and that it is broken to the  
MSSM at or near the GUT scale. We assume one of the features that remains   
is the unification of third generation Yukawa couplings. Our goal then is to  
explore the phenomenological   
implications of $t$-$b$-$\tau$ Yukawa coupling unification   
in the MSSM, subject to $SO(10)$ motivated GUT scale boundary  
conditions valid at $M_{GUT}$.  
  
Much work has already appeared on the issue of Yukawa coupling  
unification\cite{old}.   
Unification of  $t-b-\tau$ Yukawa couplings appears to require   
large values of $\tan\beta$, the ratio of Higgs field vacuum  
expectation values ({\it vevs})  
in the MSSM.   
Scenarios with large $\tan\beta$ values are technically   
natural though some degree of fine tuning may be  
required\cite{Baer:2002hf}.  
  
In models with Yukawa unification, the third generation   
SM fermion Yukawa couplings can be calculated at the weak scale, in an  
appropriate renormalization scheme.   
In this paper, we adopt the  
$\overline{DR}$ scheme, which is convenient for  
two-loop renormalization group evolution of parameters in   
supersymmetric models.  
A central feature of phenomenological analyses of SUSY models with  
Yukawa coupling unification is that weak scale supersymmetric threshold  
corrections to fermion masses (especially for the $b$-quark mass $m_b$)  
can be large\cite{hrs}, resulting in a non-trivial dependence of  
Yukawa coupling unification on the entire spectrum of SUSY particles.  
The gauge and Yukawa couplings, and the various soft SUSY breaking  
parameters, can be evolved to the grand unified scale, defined  
as the scale $Q=M_{GUT}$ at which the gauge couplings $g_1$ and $g_2$ meet.  
The coupling $g_3$ is not exactly  
equal to $g_1$ and $g_2$ at this scale. The difference is to be attributed  
to $GUT$ scale threshold corrections, whose magnitude depends  
on the details of physics at $Q = M_{GUT}$.   
At $M_{GUT}$, the third generation Yukawa   
couplings can be examined to check how well they unify.   
The measure of unification adopted in this paper is given by  
\begin{equation}  
R=\max(f_t,f_b,f_\tau)/\min(f_t,f_b,f_\tau),   
\label{Eq:DefR}  
\end{equation}  
where $f_t$, $f_b$ and $f_\tau$ are the $t$, $b$ and $\tau$ Yukawa  
couplings, and $R$ is measured at $Q=M_{GUT}$. Notice that $R$   
measures the amount of non-unification between the largest and  
the smallest of the third generation Yukawa couplings, and that the   
unified Yukawa coupling is roughly mid-way between these.  
By requiring   
unification of Yukawa couplings to a given precision, rather severe  
constraints on model parameter space can be developed.   
  
In Ref. \cite{Baer:1999mc}, it was found that Yukawa coupling unification   
to better than   
5\% ($R<1.05$) was not possible in the minimal   
supergravity (mSUGRA or CMSSM) model. This was due in part to a   
breakdown in the radiative electroweak symmetry breaking   
mechanism (REWSB) at large values of $\tan\beta$.   
However, in many $SO(10)$ models, additional  
$D$-term contributions\cite{dterms} to scalar masses are expected   
due to the reduction in   
gauge group rank upon breaking $SO(10)$ to the SM gauge symmetry.  
For $SO(10)\to SU(5)\times U(1)_X\to SU(3)_c\times SU(2)_L\times U(1)_Y$,   
the $D$-term contributions modify GUT scale scalar masses   
such that  
\begin{eqnarray*}  
m_Q^2=m_E^2=m_U^2=m_{16}^2+M_D^2 , \\  
m_D^2=m_L^2=m_{16}^2-3M_D^2 , \\  
m_N^2 = m_{16}^2+5M_D^2,\\  
m_{H_{u,d}}^2=m_{10}^2\mp 2M_D^2 ,  
\end{eqnarray*}  
where $M_D^2$ parameterizes the magnitude of the $D$-terms.  
Owing to our ignorance of the gauge symmetry breaking mechanism,  
$M_D^2$ can be taken as a free parameter,   
with either positive or negative values.  
$|M_D|$ is expected to be of order the weak scale.  
Thus, the $D$-term ($DT$) model is characterized by the following free   
parameters,  
\begin{eqnarray*}  
m_{16},\ m_{10},\ M_D^2,\ m_{1/2},\ A_0,\ \tan\beta ,\ {\rm sign}(\mu ).  
\end{eqnarray*}  
The range of $\tan\beta$ is restricted by the requirement of  
Yukawa coupling unification, and so is tightly constrained to a narrow  
range near $\tan\beta\sim 50$. For values of $M_D^2>0$, it was found  
in Ref. \cite{Baer:1999mc} that in fact Yukawa unified models with $R<1.05$   
could be generated,   
but only for values of $\mu <0$. Details of the analysis and further  
exploration of the phenomenology including  
the impact of $b\to s\gamma$ decay rate, neutralino relic density  
$\Omega_{\tz_1} h^2$ and collider search possibilities were presented in   
Ref. \cite{Baer:2000jj}. It was found that the   
branching fraction $BF(b\to s\gamma )$ was particularly constraining,  
as it is for many models with $\mu <0$ and large $\tan\beta$.  
Meanwhile, the neutralino relic density turned out acceptable over wide  
regions of parameter space due to large neutralino annihilation rates through  
$s$-channel $A$ and $H$ exchange diagrams.  
  
The precision measurement of muon $g-2$ by the E821 experiment\cite{e821}  
in 2001 provides additional  
evidence that, for supersymmetric models with   
gravity-mediated SUSY breaking, the positive sign of $\mu$ appears  
to be favored.  
After progress in the SM $(g-2)_\mu$ calculation, and further experimental  
data analysis, there is still a preference for $\mu >0$ models, but  
the latest analyses show this is somewhat weaker than first indications.   
The combination of constraints from $BF(b\to s\gamma )$ and   
$a_\mu =(g-2)_\mu$  
led various groups to   
carefully examine Yukawa unified models with  
$\mu >0$.   
  
Blazek, Dermisek and Raby (BDR)\footnote{These calculations
represent an improvement upon earlier results presented in
Ref. \cite{blazek}.} used a top-down RGE approach with exactly unified
Yukawa couplings to evaluate  
a variety of SM observables in supersymmetric models, including the  
spectrum of third generation fermions\cite{bdr}.  Performing a $\chi^2$  
analysis, they found regions of parameter space consistent with Yukawa  
coupling unification in a model with universal matter scalar masses, but  
with independent soft SUSY breaking Higgs masses (Higgs splitting or  
$HS$ model). These models were generally consistent with heavy scalar  
masses $m_{16}\sim 1.5-2.5$ TeV. In addition, a light pseudoscalar Higgs  
mass $m_A\sim 115-140$ GeV, and a small value of $\mu \sim 100-200$~GeV  
were strongly preferred. They emphasized that the $HS$ model gives a 
better fit to data than the $DT$ model.

Calculations exploring Yukawa coupling unification were also 
performed by Baer and Ferrandis (BF), but adopting a bottom-up  
approach, and focussing mainly on the $DT$ model. 
Scanning $m_{16}$ values up to 2 TeV, Yukawa  
unified solutions with $R\sim 1.3$ at best were found, but with similar  
relations amongst the $m_{16}$, $m_{10}$ and $A_0$  
parameters\cite{Baer:2001yy}. The BF analysis also examined the $HS$ 
model and found similar levels of Yukawa coupling unification.
\footnote{In Ref. \cite{nath}, 
it is shown that Yukawa unification can occur in  
models with scalar mass universality, if {\it gaugino mass  
non-universality} is allowed. In addition, Yukawa coupling quasi-unification
was explored in Ref.~\cite{pallis}.}

Both the BDR and the BF analyses found   
that the best  
solutions tended to have relations amongst boundary conditions  
\begin{equation}  
m_{10}\simeq \sqrt{2}m_{16},\ \ \ A_0\simeq -2m_{16} .  
\label{RIMH}  
\end{equation}  
These conditions were previously obtained by Bagger {\it et al.} in the  
context of radiatively driven inverted scalar mass hierarchy (RIMH)  
models\cite{jonb}. A crucial element of the RIMH solution was that third  
generation Yukawa couplings unified. As an output, it was found that
soft SUSY breaking terms
respected $SO(10)$ gauge symmetry at $M_{GUT}$.  In the RIMH model,  
multi-TeV scalar masses were adopted at the $GUT$ scale. First and  
second generation matter scalar masses remained at multi-TeV levels at  
the weak scale, thus suppressing SUSY flavor and $CP$ violating  
processes. Third generation scalar and Higgs boson masses are 
driven to TeV or below levels by their large Yukawa couplings,
so that the models may be consistent with constraints from  
fine-tuning.  
  
The RIMH models were investigated in detail in Refs. \cite{imh}, where
it was found that viable RIMH model spectra could be generated for
either sign of $\mu$.  However, adopting a complete RGE SUSY spectrum
solution including radiative electroweak symmetry breaking (REWSB) along
with a realistic value of $m_t$, the magnitude of the hierarchy between
first and third generation sfermion masses was found to be more limited
than what is suggested by the approximate analytic solution of
Ref. \cite{jonb}.  A marginally larger hierarchy could be obtained for
the $HS$ model as compared to $DT$ model.

In this paper, we perform an updated analysis of third generation   
Yukawa coupling unification in supersymmetric models.  
Our motivation for an updated analysis is as follows.
\begin{itemize}  
\item   
We now incorporate the complete one-loop   
self energy corrections to fermion masses\cite{pierce}, whereas previously  
we used approximate formulae. We also include an improved   
treatment of running parameters in the self-energy calculation, and have  
also corrected a bug in our evaluation of the Passarino-Veltman $B_1$  
function present in ISAJET versions prior to 7.64. These last two  
improvements give good agreement between ISAJET\cite{isajet} and also the   
Suspect\cite{suspect}, SoftSUSY\cite{softsusy} and   
Spheno\cite{spheno} Yukawa coupling calculations,   
as extracted by Kraml\cite{kraml}.  
The main effect of the improved self energy treatment is an extension of the  
allowed SUSY model parameter space to much larger values of GUT scale  
scalar masses. The boundary of this region, determined by  
radiative electroweak symmetry breaking (REWSB) constraints,   
is very sensitive in   
particular to the value of the top quark Yukawa coupling, and at large   
$\tan\beta$, also to the $b$-quark Yukawa coupling.  
  
\item   
 We have also included an   
improved treatment of the value of $m_b^{\overline{DR}}(M_Z)$.  
The two-loop analysis  
of Ref. \cite{Baer:2002ek} yields $m_b^{\overline{DR}}(M_Z) =2.83\pm  
0.20$ GeV.\footnote{In quoting this error, we conservatively use the  
Particle Data Group value for $m_b^{\overline{DR}}(m_b)$. More recent extractions of this  
suggest that the error on $m_b^{\overline{DR}}(M_Z)$ may be closer to  
$\pm 0.1$~GeV.}  
The central value of $m_b^{\overline{DR}}(M_Z)$ serves as one of the boundary   
conditions for the  RG analysis of Yukawa coupling unification.  
  
\item We include analyses of the neutralino relic density $\Omega_{\tz_1} h^2$,  
$BF(b\to s\gamma )$ decay, $a_\mu$ and $BF(B_s\to \mu^+\mu^- )$.  
Comparison of model predictions for these with corresponding experimental   
measurements (or experimental limits) yields significant  
constraints on model parameter space.  
  
\item We expand the parameter space over which our model scans take place.  
This turns out to be important especially for $\mu >0$ models, where we  
find that {\it Yukawa coupling unification to better than 1\% can occur,  
but only  
at very high values of} $m_{16}\sim 8-20$ TeV!  
\end{itemize}  
  
The rest of this paper is organized as follows. In Sec. 2, we present  
calculations of Yukawa coupling unification for supersymmetric models  
with $\mu >0$. In the mSUGRA model, we find Yukawa coupling   
unification only to 35\% is possible.  
In contrast, in the $DT$ model, Yukawa unification   
down to about 10\% is possible,   
where the best unification occurs for  
very large values of $m_{16}\agt 10$ TeV, while $A_0\simeq -2m_{16}$.  
For the $HS$ model, we find that   
Yukawa unification to less than 1\% is possible, but only for {\it very}  
large values of GUT scale scalar masses $m_{16}\sim 8-20$ TeV,   
with low values of $m_{1/2}\lesim 300-400$ GeV, and using RIMH  
boundary conditions~(\ref{RIMH}).   
For the $HS$ models with good Yukawa coupling   
unification, we find good agreement in general with constraints from  
$b\to s\gamma $ and $(g-2)_\mu$. However, it appears difficult to  
achieve a reasonable value of the neutralino relic density   
$\Omega_{\tz_1}h^2$ unless  
the parameter $m_{16}$ is less than typically a 
few TeV.
In addition, if $m_{16}\sim 8-20$ TeV, then third generation scalars  
occupy mass ranges of typically 2-10 TeV, 
which can cause some further tension  
if fine-tuning constraints are adopted.  
In Sec. 3 we present updated calculations for $\mu <0$. In the mSUGRA model,  
we find perfect Yukawa unification is possible, but only for   
extreme parameter values such as $m_0\sim 10$ TeV with   
$m_{1/2}\sim 15$ TeV.   
However, in  
both the $DT$ and $HS$ models, perfect Yukawa unification can be achieved  
for $m_{1/2}$ values as low as $1-2$ TeV, which should be consistent  
with fine-tuning constraints.   
These models offer  
regions of parameter space consistent with the neutralino relic density,   
but have tension with $b\to s\gamma$ and $(g-2)_\mu$ unless  
$m_{1/2}\agt 2-3$ TeV.  
In Sec. 4, we compare our approach with that of
BDR, and present  
some additional calculational details. In Sec. 5,   
we present our conclusions.

\section{Supersymmetric models with $\mu >0$}  
  
\subsection{mSUGRA model}  
  
To establish a baseline for our studies of Yukawa coupling   
unification, we first examine the extent to which Yukawa coupling   
unification occurs in the paradigm minimal supergravity model\cite{msugra}  
(mSUGRA, or CMSSM model). The parameter space is given by  
\begin{equation}  
m_0,\ m_{1/2},\ A_0,\ \tan\beta ,\ {\rm and}\ {\rm sign}(\mu ).  
\end{equation}  
We take the pole mass to be $m_t=175$ GeV.  
We then perform the simple exercise of scanning the mSUGRA  
model parameter space for $\mu >0$ over the range  
\begin{eqnarray*}  
0< m_0< 5\ {\rm TeV},\\  
0< m_{1/2}< 5\ {\rm TeV},\\  
-7\ {\rm TeV} < A_0 < 7\ {\rm TeV},\\  
2 < \tan\beta < 60   
\end{eqnarray*}  
using the ISAJET v7.64 program.  
  
The results of the scan are shown in Fig. \ref{fig:sugp}, where we   
display only solutions satisfying $R<1.6$. Solutions which are valid  
with the exception of being excluded by LEP2 constraints   
($m_{\tw_1} > 103$~GeV, $m_{\te_R} > 99$~GeV if $m_{\tell_R}-m_{\tz_1}>10$ GeV,
$m_{\ttau_1}>76$ GeV and   
$m_h > 114$~GeV) are shown as crosses,  
while solutions in accord with LEP2 searches are denoted by dots.  
We see immediately that the $t$-$b$-$\tau$ Yukawa unification  
reaches the 35\% level at best. The best unified solutions occur  
for large $m_0\simeq 3-5$ TeV, low $m_{1/2}\lesim 0.5$ TeV,   
$\tan\beta \sim 50-55$, and $A_0\lesim 3-4$ TeV.   
Additional scans with $m_0$ up to 10 TeV gave no further  
improvement upon Yukawa unification.  
It is noteworthy that  
if we restrict solutions to the range $m_0\alt 1$ TeV, then the Yukawa  
coupling unification becomes considerably worse.   
To improve upon the situation, we examine models with non-universal  
scalar masses.  
\FIGURE[t]{  
\epsfysize=14cm\epsfbox[0 25 567 510]{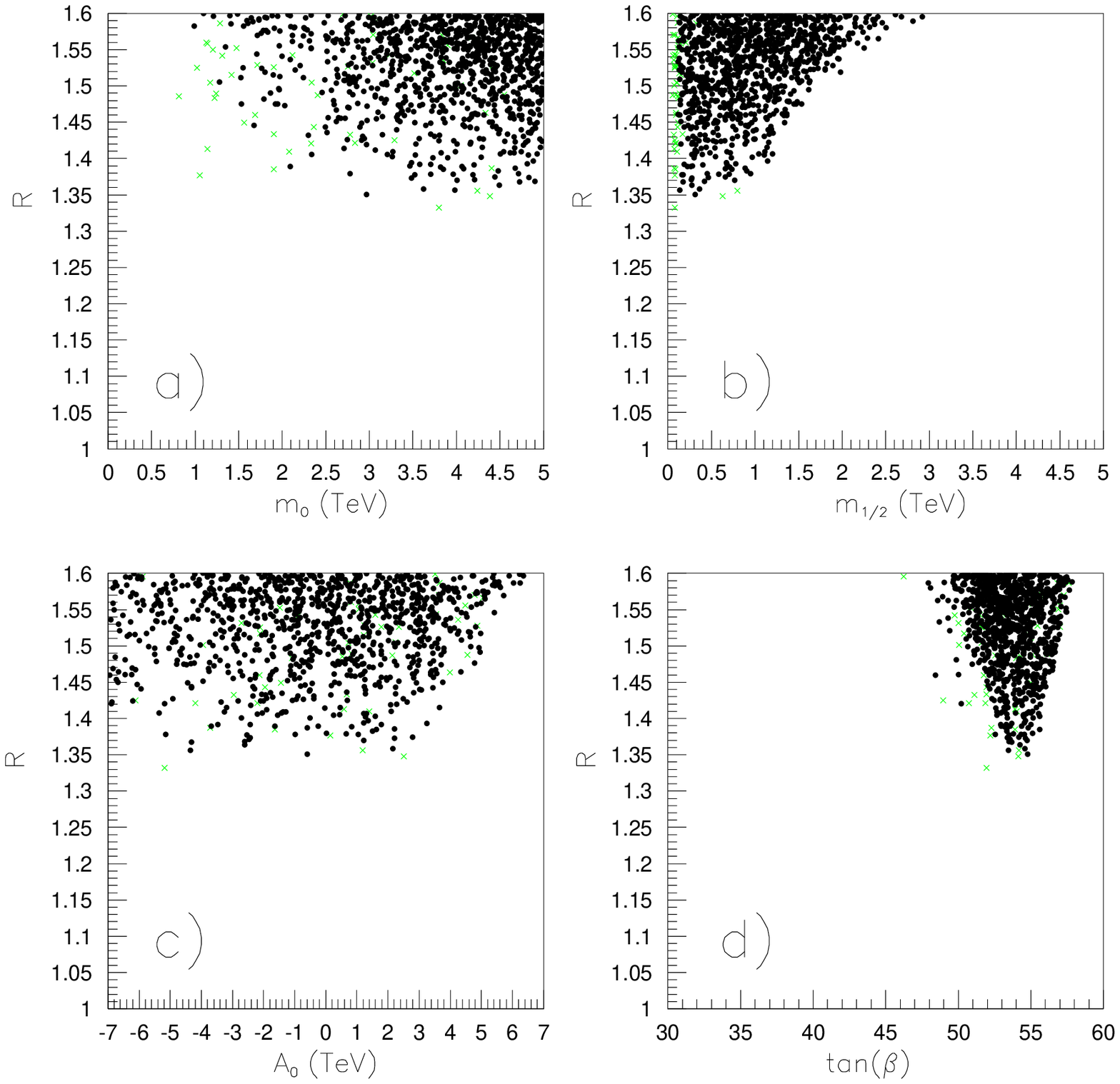}  
\caption{Plot of Yukawa unification parameter $R$ versus  
input parameters of the mSUGRA model for {\it a}) $m_0$,   
{\it b}) $m_{1/2}$, {\it c}) $A_0$ and {\it d}) $\tan\beta$,  
when $\mu >0$. Points denoted by crosses are allowed solutions that  
are excluded by LEP2 constraints mentioned in the text.}  
\label{fig:sugp}}  

\subsection{$DT$ model}  
  
First, we re-assess Yukawa coupling unification for $\mu >0$ within the  
$DT$ model. We scan this model over the following parameter range:  
\begin{eqnarray}  
0&<&m_{16}<20\ {\rm TeV}, \nonumber\\  
0&<&m_{10}<30\ {\rm TeV}, \nonumber\\  
0&<&m_{1/2}<5\ {\rm TeV}, \nonumber\\  
-(m_{10}/\sqrt{2})^2&<&M_D^2<+(m_{10}/\sqrt{2})^2,\label{rangesI}\\  
40&<&\tan\beta <60,       \nonumber\\  
-3m_{16}&<& A_0<3m_{16}.\nonumber  
\end{eqnarray}  
Thus, the parameter scan range is greatly increased over the earlier BF  
analysis\cite{Baer:2001yy}. Our results are shown in  
Fig. \ref{fig:DTmupdots}, where we plot the resulting $R$ value against  
various parameter and ratio of parameter choices.  Models marked by dark  
blue dots are results of a wide random scan of the full parameter region  
indicated by Eq.(\ref{rangesI}). Results of a dedicated narrow scan are  
shown in light blue.  
  
\FIGURE[t]{\hspace*{-.5cm}\epsfig{file=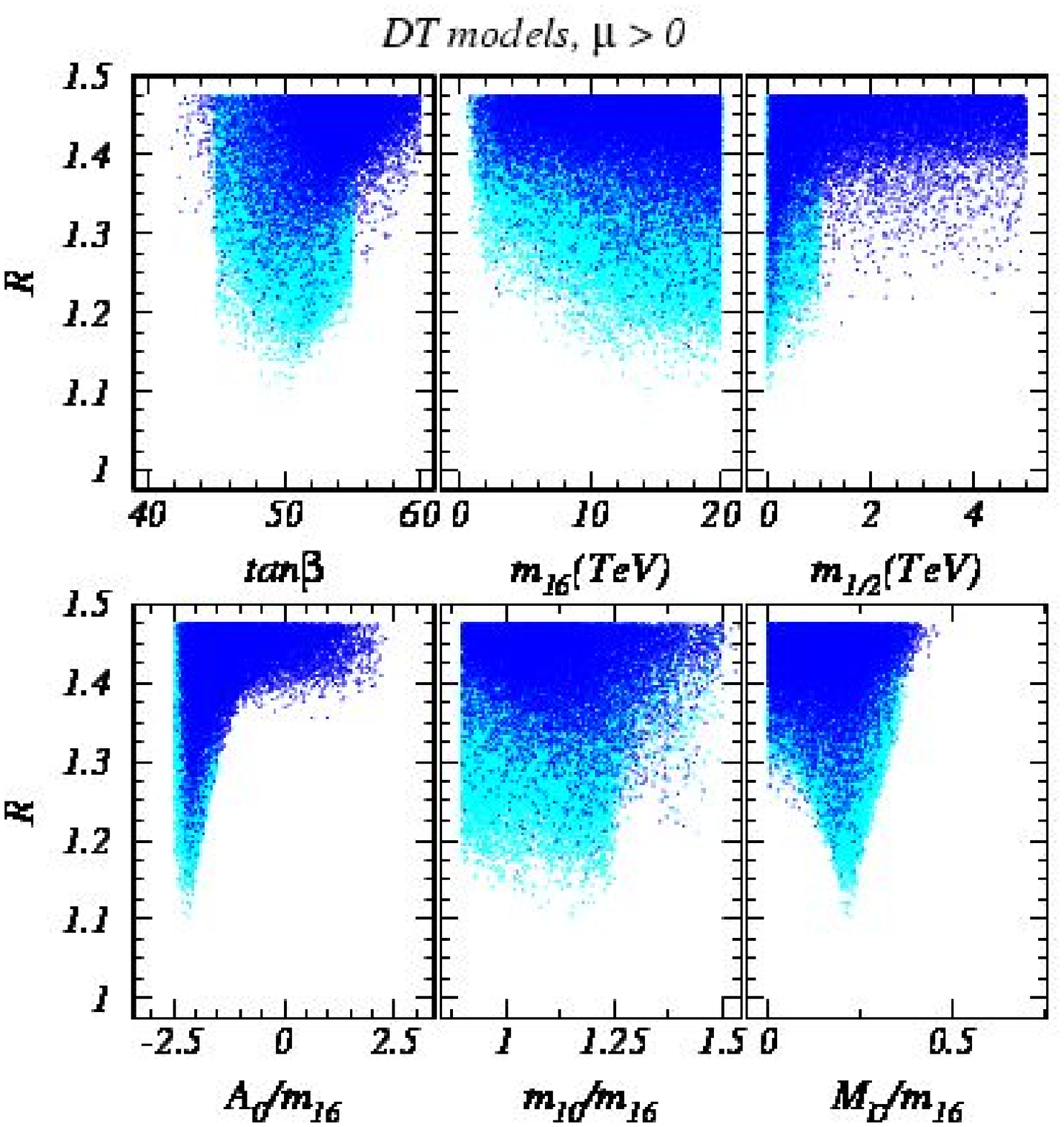,height=15cm}  
\caption[]{\label{fig:DTmupdots}  
Plot of $R$ versus parameters of the $DT$ model with $\mu >0$   
for  
{\it a}) $\tan\beta$,   
{\it b}) $m_{16}$, {\it c}) $m_{1/2}$, {\it d}) $A_0/m_{16}$,   
{\it e}) $m_{10}/m_{16}$ and {\it f}) ${\rm sign}(M_D^2)\sqrt{|M_D^2|}/m_{16}$.}}  
  
We see that the best   
Yukawa unification possible gives $R\simeq 1.1$, {\it i.e.}   
Yukawa unification to 10\% (really $\pm 5$\%). The first three   
frames show that the best unification occurs  
for $\tan\beta \sim 50$, and for $m_{16}\agt 10$~TeV, and small values  
of  
$m_{1/2}\sim 0-0.5$ TeV. We have checked that if we restrict the parameter  
range to $m_{16} \alt 2$ TeV, then $R$ is always larger than 1.25,  
which is close to the result obtained in   
Ref. \cite{Baer:2001yy}. The fourth frame shows $R$ versus the ratio  
$A_0/m_{16}$. Here we see that the best Yukawa unification occurs  
sharply next to $A_0\sim -2m_{16}$, as in the RIMH scenario, and as in the  
previous BF analysis. The fifth frame shows $R$ versus the ratio  
$m_{10}/m_{16}$. Here, the minimum occurs near $m_{10}\sim 1.2 m_{16}$,  
{\it i.e.} close to but   
somewhat below the optimal RIMH value of $m_{10}=\sqrt{2}m_{16}$.  
Finally, we show $R$ versus the ratio $sign(M_D^2)\sqrt{|M_D^2|}/m_{16}$.   
In this case, the best Yukawa unification occurs at $M_D\sim 0.25 m_{16}$.  
Choosing $M_D=0$ brings us almost back the mSUGRA case (except that  
$m_{10}$ need not equal $m_{16}$), and the best unification is just  
$R_{min}=1.28$.   
  
In addition to performing random scans over parameter space, we also  
scanned for the minimum of $R$ using the MINUIT minimization   
program. In this case, similar results were obtained, with the  
minimum occurring for large values   
of $m_{16}\sim 15$ TeV, and $m_{1/2}\sim 0$.\footnote{Note that  
$m_{1/2}$ values near zero can be allowed in our analysis and be consistent  
with LEP2 constraints on the chargino mass because the gaugino mass  
RGEs receive significant two-loop contributions owing to the very large  
values of soft SUSY breaking parameters.}   
  
\subsection{$HS$ model}  
  
Next, we turn to the $HS$ model. BDR have pointed out that threshold  
corrections due to the third generation right-handed neutrino (which  
couples to $H_u$ but not $H_d$) naturally lead to non-degenerate
$m_{H_u}^2$ and $m_{H_d}^2$. As in the $DT$ model, 
the Higgs mass splitting may facilitate REWSB. 
The difference is that squark and slepton mass parameters are  
unaffected in the $HS$ model.   
We adopt the same parameter space  
as in the $DT$ model, except that this time the splitting  
applies only to the two Higgs multiplets of the MSSM, while all  
matter scalars have a common GUT scale mass value given by $m_{16}$.  
  
\FIGURE[t]{\hspace*{-.5cm}\epsfig{file=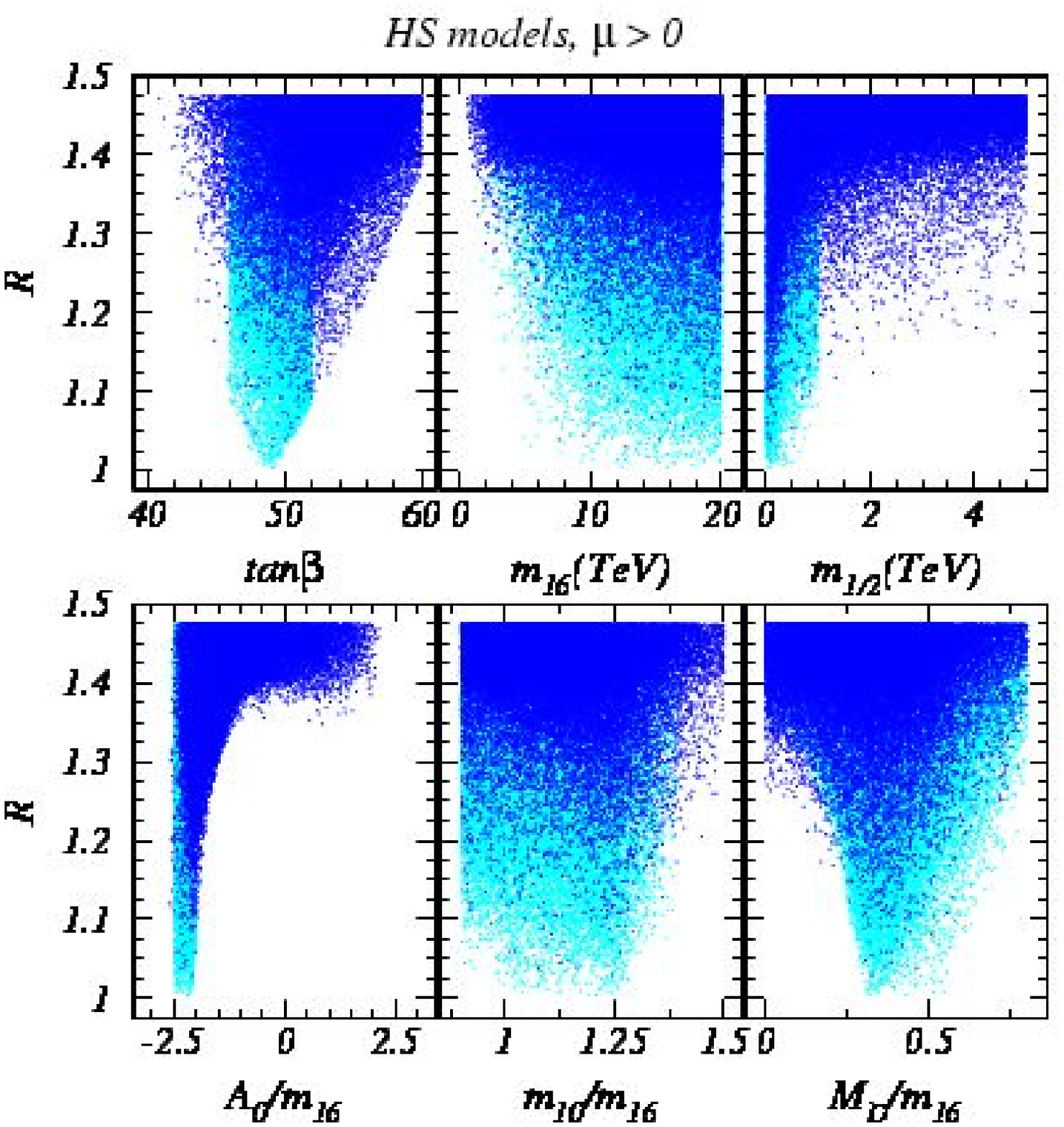,height=15cm}  
\caption[]{\label{fig:HSmupdots}  
Plot of $R$ versus parameters of the $HS$ model with $\mu >0$   
for  
{\it a}) $\tan\beta$,   
{\it b}) $m_{16}$, {\it c}) $m_{1/2}$, {\it d}) $A_0/m_{16}$,   
{\it e}) $m_{10}/m_{16}$ and {\it f}) ${\rm sign}(M_D^2)\sqrt{|M_D^2|}/m_{16}$.}}  
  
We scan over the same parameter ranges as in the $DT$ model case,  
with the results shown in Fig.~\ref{fig:HSmupdots} by the dark blue dots.  
Corresponding results  
of a more focussed scan over a restricted but optimized range of  
parameters are shown by the light blue dots. The most noteworthy  
feature is that we find essentially exact Yukawa coupling unification  
is possible. The Yukawa unified models are characterized by  
parameter choices of $\tan\beta\sim 49$, $m_{16}\agt 8$~TeV and  
$m_{1/2}\sim 0-0.4$ TeV, as shown by the first three frames of the  
figure.  
  
The fourth frame of Fig.~\ref{fig:HSmupdots} shows $R$ versus $A_0/m_{16}$,  
and illustrates the sharp minimum of Yukawa unified models at   
$A_0\sim -(2- 2.5)m_{16}$. A choice of $A_0 \alt -2.5m_{16}$ leads  
to tachyonic masses, while $A_0 \agt -2m_{16}$  
results in much less unified values of Yukawa couplings.   
The fifth frame shows that  
Yukawa unification occurs for $m_{10}\sim (1-1.3) m_{16}$, again somewhat lower  
than the RIMH optimal choice of $m_{10}\simeq \sqrt{2}m_{16}$.  
The last frame shows $R$ versus ${\rm sign}(M_D^2)\sqrt{|M_D^2|}/m_{16}$.  
Here, a somewhat bigger range of $M_D^2$ (compared to the  
$DT$ model) yields Yukawa unified solutions which are now obtained for  
$0.25 \lesssim M_D/m_{16} \lesssim 0.5$.   
Reducing the value of $M_D$ to zero only allows,  
at best, solutions with $R$ down to 1.28.  
In contrast to earlier work on RIMH models,  
our improved treatment of fermion self-energies in this analysis allows us  
to find Yukawa unified solutions by accessing much larger values of $m_{16}$  
that previously resulted in a breakdown of the REWSB mechanism\cite{imh}.   
We note here that we have performed similar scans using values of
$m_t=172$ and $180$ GeV (with $m_b^{\overline{DR}}(M_Z)=2.83$ GeV), 
and $m_b^{\overline{DR}}(M_Z)=3.03$ GeV (with $m_t=175$ GeV), and in each 
case have found results qualitatively similar to those shown in
Fig. \ref{fig:HSmupdots}: in particular, $R \simeq 1$ is only obtained
for $m_{16} \agt 8-10$~TeV

%
%
  
Our next results are shown in  
Fig. \ref{HSplane_1}. Here, we show the $m_{16}\ vs.\ m_{1/2}$ plane for  
$m_{10}=1.24 m_{16}$, $A_0=-2.05 m_{16}$, $\tan\beta =49.7$ and $\mu >0$.  
The mass splitting applied {\it only} to the Higgs multiplets is  
parametrized by $M_D=0.334 m_{16}$.   
The contours show regions where $R<1.03$ and $R<1.02$, {\it i.e.} Yukawa  
unification to better than 2-3\%. As seen from the figure, 
these regions occur  
at extremely large values of $m_{16}\simeq 8-20$ TeV.   
Since scalar masses are so large, it is critical  
to perform coupling constant {\it and} soft term mass evolution  
using two-loop RGEs\cite{mv}. The resulting   
first and second generation multi-TeV scalar masses are sufficient  
to suppress most SUSY flavor and $CP$ violating processes, and offers  
a decoupling solution to the SUSY flavor and $CP$ problems.
This comes at some fine-tuning expense, since third generation  
scalars, while suppressed, typically have masses in the few TeV range.

\FIGURE[t]{\hspace*{-.5cm}\epsfig{file=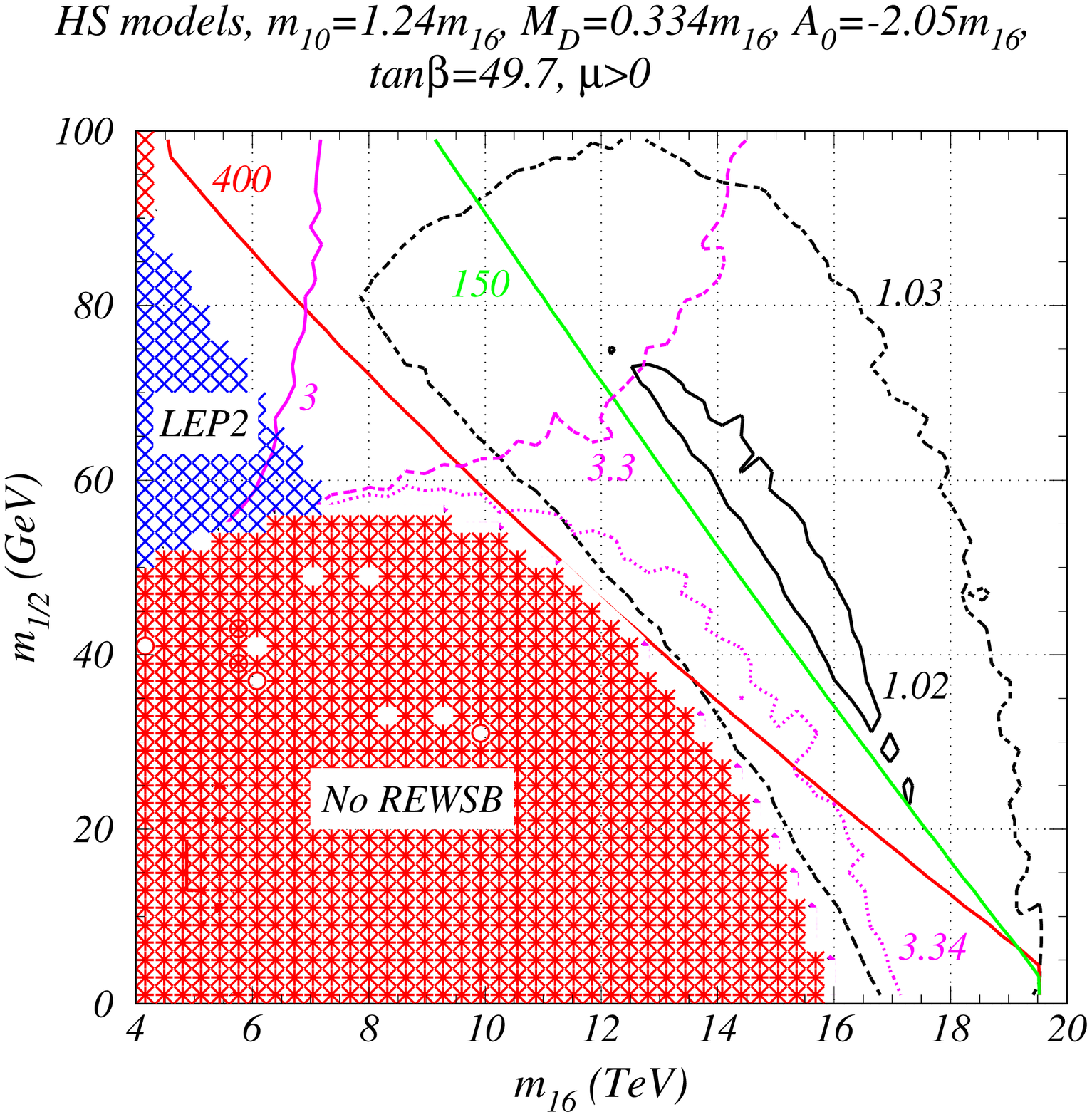,height=16cm}  
\caption[]{\label{HSplane_1}  
A plot in the $m_{16}\ vs.\ m_{1/2}$ plane for $\tan\beta =49.7$, showing   
contours of Yukawa unification parameter $R$ (black), contours of $BF(b\to   
s\gamma)$ ($\times 10^4$) (magenta), and contours of $m_{\tg}=400$ GeV (red)   
and $m_{\tw_1}=150$ GeV (green).}}  
%
  
  
A characteristic feature of these solutions is that $m_{1/2}$ is   
rather small, typically less than $150$ GeV, giving rise to  
relatively light masses for the $\tg$ and lighter charginos and   
neutralinos.   
Also shown in Fig. \ref{HSplane_1} are contours of $m_{\tg}=400$ GeV and  
$m_{\tw_1}=150$ GeV. The former may just be accessible  
to Fermilab Tevatron searches for $\tg\tg$ production  
with large values of $m_{\tq}$\cite{tevgg},   
while the latter  
may be accessible via $\tw_1\tz_2\to 3\ell$   
searches\cite{tev3l,sugrarep}.  
  
Representative examples   
of sample solutions are shown in Table \ref{tab:one}, with  
$m_{16}$ values of $2.5,\ 5.31,\ 10$ and $14$ TeV, and also a small
$\mu$ point with $m_{16}=1500$~GeV (pt. 5) that we will discuss
separately. We list the  
GUT scale values of the three Yukawa couplings, plus a variety  
of physical SUSY particle masses. In particular, $m_{\tst_1}$  
varies from $0.24-3.3$ TeV for these solutions.   
We also list the ``crunch factor'' $S$ as defined in  
Ref. \cite{jonb,imh}:   
\begin{eqnarray}  
S   =  
 {{3(m_{\tu_L}^2+m_{\td_L}^2+m_{\tu_R}^2+m_{\td_R}^2)+m_{\te_L}^2+m_{\te_R}^2+  
m_{\tnu_e}^2}  
 \over  
{3(m_{\tst_1}^2+m_{\tb_1}^2+m_{\tst_2}^2+m_{\tb_2}^2)+m_{\ttau_1}^2+  
m_{\ttau_2}^2+ m_{\tnu_{\tau}}^2}}.  
\label{crunch}  
\end{eqnarray}  
We see that the solutions typically have $S\sim 6-7$, in accord with  
previous results from Ref. \cite{imh}, but lower than those generated  
by the approximate analytic calculations in Ref. \cite{jonb}.  

Pt.~5 in Table~\ref{tab:one} has been chosen to try to
duplicate the first 
point in Table~1 of BDR, and to show that we are able
to find solutions with a small values of $\mu$. The only difference in 
input parameters from BDR is that we take $m_{10}/m_{16}=1.315$ instead of 
1.35, so that REWSB is satisfied in ISAJET. 
We differ from BDR in that for solutions with small $\mu$,
we are unable to obtain $R$ smaller than about 1.2. Pt. 5 (or
a variant thereof) which is unequivocally excluded by the very large value for
$BF(b \to s\gamma)$, would otherwise be quite acceptable on
phenomenological grounds. The higgsino component of the neutralino
allows efficient annihilation of cosmological LSPs, resulting in too low
a relic density to account for all dark matter. The value of $BF(B_s
\to \mu^+\mu^-)$ appears to be within reach of Run 2 of the
Tevatron. Indeed the Tevatron may be able to directly discover charginos
and neutralinos via trilepton searches, while a plethora of signals
should be observable at the LHC. However, since this point is excluded
we will not refer to it any further.

\begin{table}  
\begin{center}  
\caption{\label{tab:one}Model parameters, Yukawa couplings and corresponding  
sparticle masses for five case studies in the $HS$ model with $\mu >0$.}  
\bigskip  
\begin{tabular}{lccccc}  
\hline  
parameter                          	& pt. 1 &  pt. 2 &  pt. 3 &
pt. 4 & pt. 5\\  
\hline  
$m_{16}$                        	& 2493.0& 5310.0 & 10000.0&
14000.0 & 1500\\  
$m_{10}$                        	& 3226.0& 6595.0 & 12420.0&
17360.0 & 1972.5\\  
$M_D   $                        	& 975.0 & 1778.8 &  3350.0&
4676.0 & 502.9\\  
$m_{1/2}$                       	& 130.0 &   79.0 &    79.0&
65.0 & 250 \\  
$A_0$
&-4754.0&-10885.5&-20500.0&-28700.0 & -2745.0 \\  
$\tan\beta$                     	& 50.6  &   49.7 &    49.7&
49.7 & 51.2\\  
$f_t(M_{GUT})$                  	& 0.507 &  0.537 &  0.557 &
0.560 & .507\\  
$f_b(M_{GUT})$                  	& 0.453 &  0.536 &  0.559 &
0.560 & 0.455 \\  
$f_\tau (M_{GUT})$              	& 0.539 &  0.561 &  0.571 &
0.571 & 0.558 \\  
$\epsilon_3=\frac{\alpha_3(M_{GUT})-\alpha_{GUT})}{\alpha_{GUT}}$
& -0.034 & -0.025 & -0.019 & -0.015 & -0.042 \\
$R$                             	&  1.19 &   1.05 &   1.03 &
1.02 & 1.23\\  
$m_{\tg}$                       	& 434.3 &  366.4 &  456.0 &
486.1 & 681.2\\  
$m_{\tu_L}$                     	& 2462.7& 5225.2 & 9849.8 &
13796.0 & 1554.1 \\  
$m_{\td_R}$                     	& 2479.3& 5255.8 & 9905.0 &
13871.6 & 1557.1\\  
$m_{\tst_1}$                    	&  418.4& 1042.1 & 2244.0 &
3335.8 & 235.1 \\  
$m_{\tb_1}$                     	&  820.2& 1520.2 & 2946.4 &
4274.2 & 587.9 \\  
$m_{\tell_L}$                   	& 2463.6& 5257.7 & 9904.1 &
13868.0 & 1494.2 \\  
$m_{\tell_R}$                   	& 2463.6& 5358.8 &10090.1 &
14124.1 & 1517.6 \\  
$m_{\tnu_{e}}$                  	& 2527.2& 5257.1 & 9903.7 &
13867.7 & 1492.0 \\  
$m_{\ttau_1}$                   	&  909.0& 2095.8 & 4069.0 &
5790.1 & 498.9 \\  
$m_{\tnu_{\tau}}$               	& 1825.8& 3960.1 & 7496.0 &
10524.0 & 1104.1 \\  
$m_{\tw_1}$                     	&  110.9&  104.1 &  141.0 &
159.9 & 129.7 \\  
$m_{\tz_2}$                     	&  110.9&  104.0 &  140.9 &
159.7 & 140.2 \\  
$m_{\tz_1}$                     	&  58.4 &   49.3 &   64.8 &
71.7 & 90.3 \\  
$m_h      $                     	& 125.6 &  127.9 &  130.7 &
131.4 & 121.0 \\  
$m_A      $                     	& 1168.6& 1018.0 & 1841.2 &
2733.2 & 525.7 \\  
$m_{H^+}  $                     	& 1174.1& 1024.1 & 1845.5 &
2737.1 & 536.3\\  
$\mu      $                     	&  303.8& 1404.5 & 2533.7 &
3480.3 & 160.5\\  
$S        $                     	&  5.94 &   6.53 &   6.37 &
6.16 & 5.26\\  
$a_\mu \times 10^{10}$          	&  3.81 &  0.362 & 0.0841 &
0.0373 & 11.05 \\  
$BF(b\to s\gamma)\times 10^4$   	&  2.83 &   2.54 &   3.23 &
3.32 & 70.9 \\  
$BF(B_s\to\mu^+\mu^-)\times 10^8$       &  1.37 &   2.89 &  0.937 &
0.610 & 11.5 \\  
$\Omega_{\tz_1}h^2$             	&  0.178&   180. &   50.4 &
32900 & 0.054 \\  
\hline  
\end{tabular}  
\end{center}  
\end{table}  

In Fig. \ref{HSevolve}, we show the evolution of gauge and Yukawa   
couplings (upper frame), and soft SUSY breaking parameters (lower frame),  
versus the renormalization scale $Q$. The example point has   
$m_{16}=10$ TeV, and corresponds to pt. 3 in Table \ref{tab:one}. 
The separate gauge and Yukawa  
coupling unifications at $Q=M_{GUT}$ are evident in the upper  
frame. In the lower frame, the  evolution characteristic to the   
RIMH framework  
shows the suppression of third generation and Higgs soft   
breaking parameters  
relative to those of the first or second generation.  

\FIGURE[t]{\hspace*{-1cm}  
\epsfysize=17cm\epsfbox[0 30 350 500]{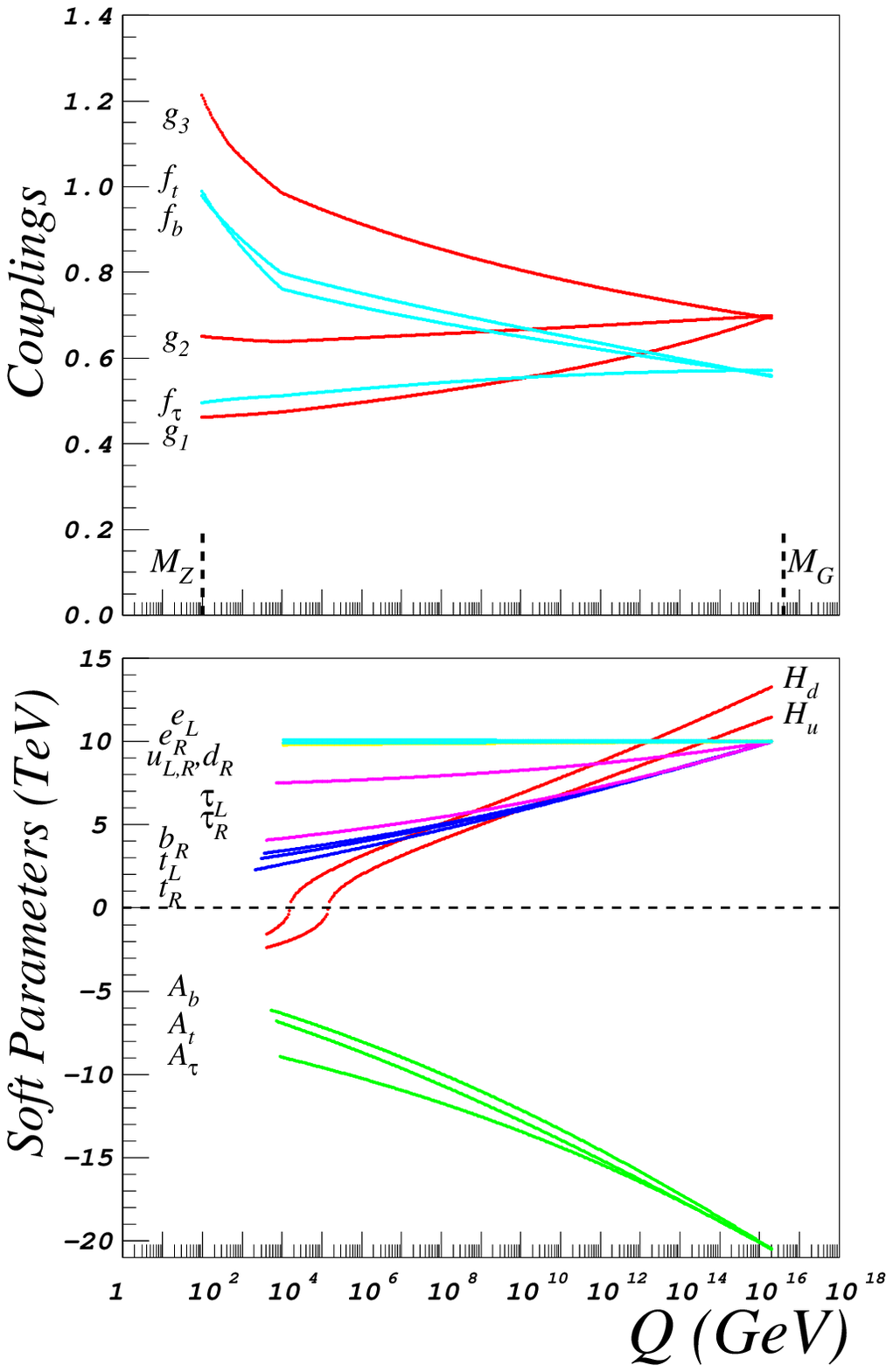}  
\caption[]{\label{HSevolve}  
Gauge and Yukawa coupling evolution (upper frame)   
and the evolution of select soft SUSY breaking parameters  
(lower frame) versus scale $Q$ for the case study pt. 3 with $m_{16}=10$  
TeV. The unification of the three gauge couplings, and  
independently,  
of the three Yukawa couplings is illustrated.  
}}  

In Fig. \ref{HSmasses}, we show selected sparticle mass spectra  
versus $m_{16}$ obtained for $\tan\beta =50.6$, $m_{1/2}=100$ GeV,   
$m_{10}=1.24m_{16}$, $M_D=0.321 m_{16}$ and $A_0=-2m_{16}$.  
The considerable gap between first and third generation sparticle  
masses is prominently displayed. As already discussed, the gap has a  
dynamical origin, and may solve the SUSY flavor and $CP$ problems
via a decoupling solution while maintaining naturalness.
\FIGURE[t]{\hspace*{-.5cm}\epsfig{file=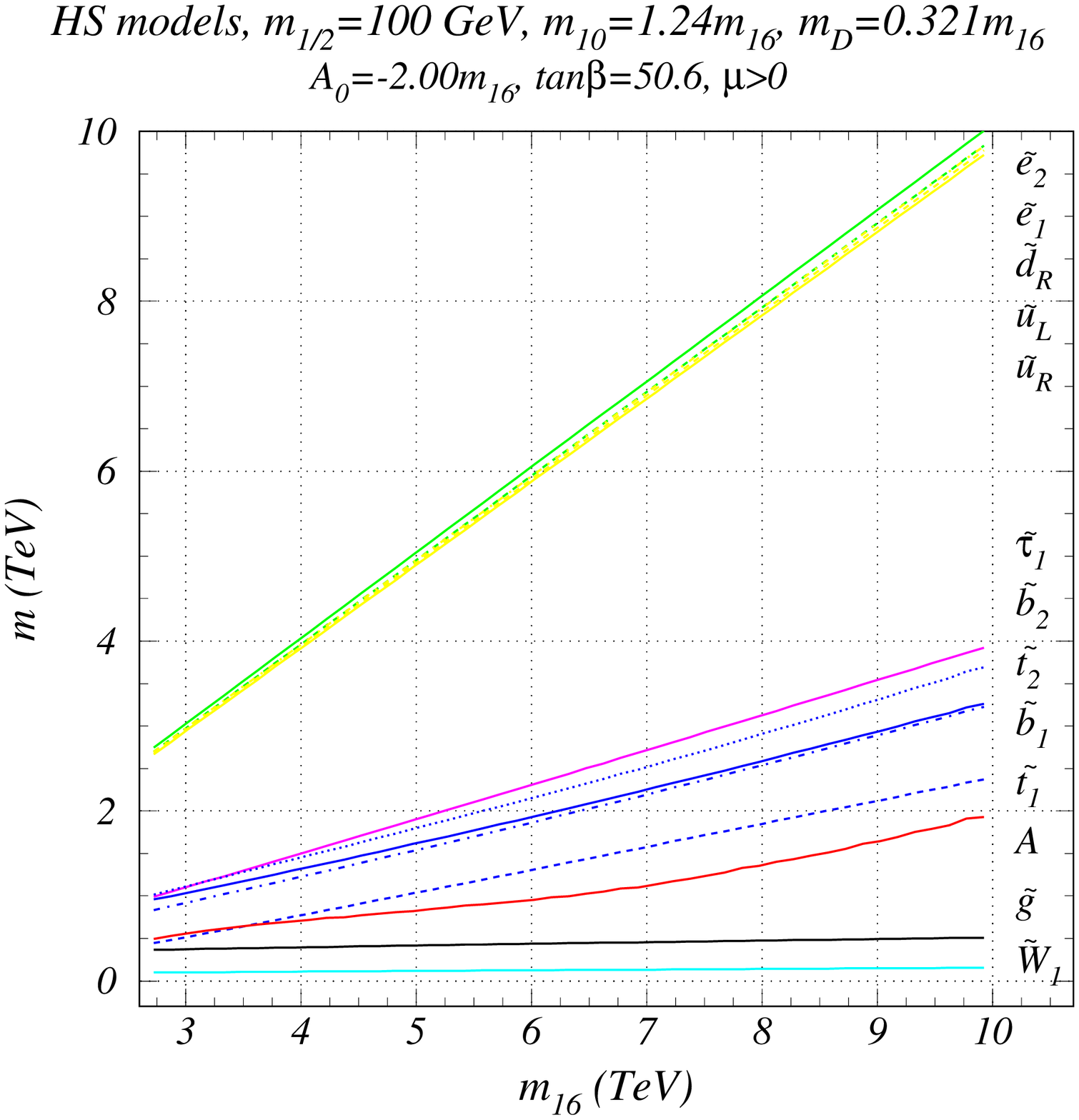,height=16cm}  
\caption[]{\label{HSmasses}  
Selected sparticle masses versus $m_{16}$ for $m_{1/2}=100$ GeV  
and other parameters as listed on the figure.}}  

In Table \ref{tab:one},   
we also list the corresponding values of $a_\mu$\cite{bbft},   
$BF(b\to s\gamma )$\cite{bsg},    
$BF(B_s\to \mu^+\mu^- )$\cite{mtw} and   
$\Omega_{\tz_1} h^2$\cite{bbb}. Many authors have evaluated these  
quantities within SUSY models: above, we cite the paper whose calculation  
we have used to obtain the results shown here.   
Contours for these observables are also shown in Fig.~\ref{HSplane_2},  
for the value of $\tan\beta =50.6$.\footnote{If model parameters 
are allowed to vary, then larger regions with a good fit to
$BF(b\to s\gamma )$ and $a_\mu$ will be allowed: see Ref.~\cite{blazek2}.}
The color characterizes the  
observable, and for each observable, the contour type (solid or  
dashed) characterizes a value for the observable. Also shown are   
regions excluded by various theoretical (no or improper  
REWSB) and experimental constraints from LEP2. In the region labelled,  
``No RGE convergence'', the solutions to the renormalization group equation  
do not satisfy the convergence criteria\footnote{Soft SUSY breaking
mass parameters for matter scalars are required to change by 
less than 0.3\% between the final iterations of the RGEs. 
For Higgs masses and the $B$ and  
$\mu$ parameters, the restriction is less restrictive by about an order  
of magnitude.} required in ISAJET.   
The value of  
$BF(b\to s\gamma )$ is below the averaged measured value of  
$BF(b\to s\gamma )=(3.25\pm 0.37)\times 10^{-4}$   
for pt. 2, but   
in closer accord with  
experiment for pts. 1, 3 and 4; however, the prediction for the first four  
points is compatible with the experimental value once theoretical  
unceertainties are taken into account.   
This is unlike the case of Yukawa unified  
models with $\mu <0$, which typically yield too large a value of   
$BF(b\to s\gamma )$. The value of $a_\mu$ is rather small for all points  
due to suppression from the multi-TeV smuon and   
sneutrino masses, but   
is not inconsistent  
with measurements from experiment $E821$. Also, the CDF collaboration has  
found $BF(B_s\to\mu^+\mu^- )<2.6\times 10^{-6}$, and expects to probe to  
about $10^{-7}$ with Run 2 data.   
Points 1-4 are likely out-of-reach for the Tevatron $BF(B_S\to\mu^+\mu^- )$   
measurements, but they will be probed at the LHC  
after about two years   
of operation.   
%
\FIGURE[t]{\hspace*{-1cm}\epsfig{file=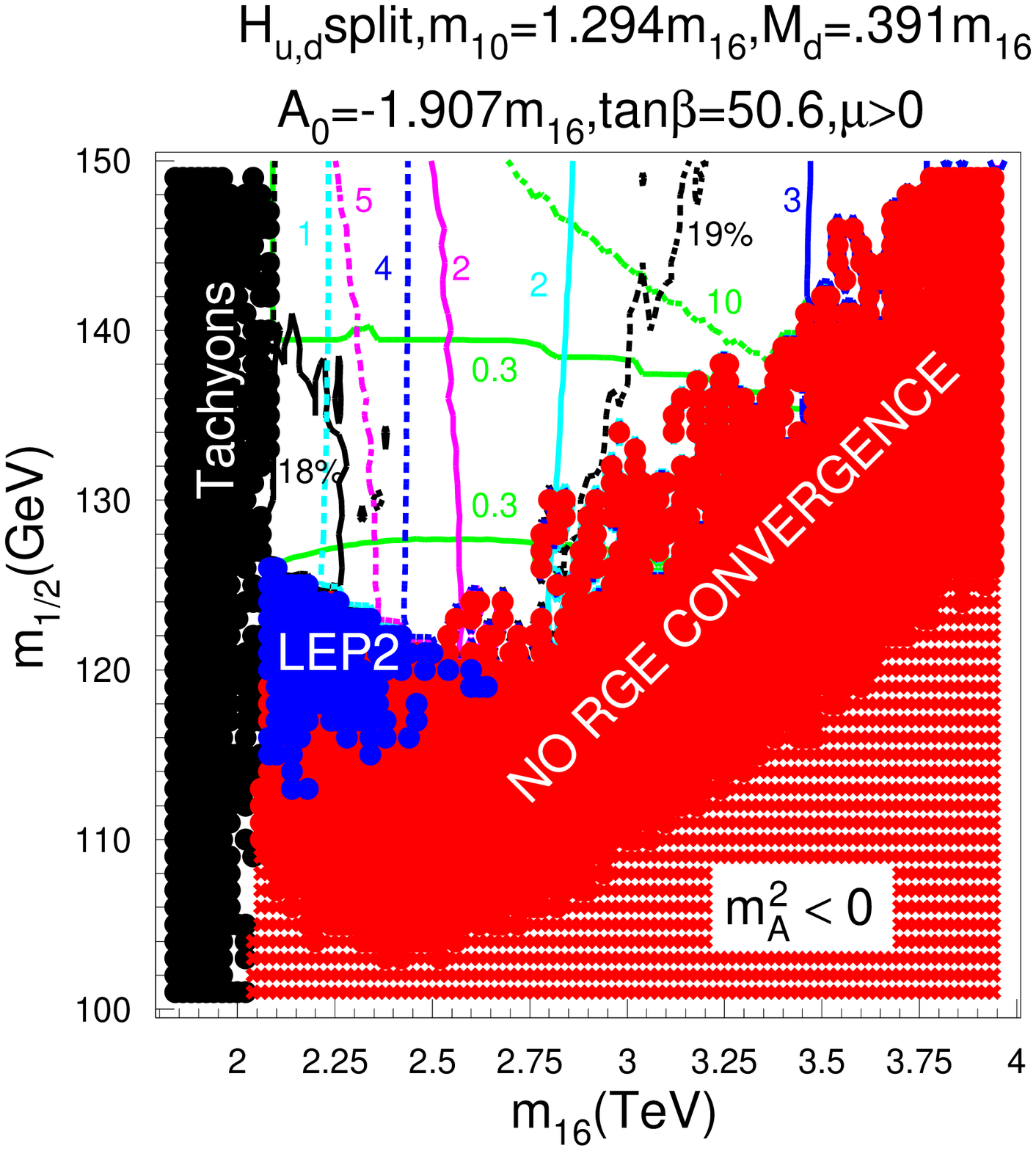,height=17cm}  
\caption[]{\label{HSplane_2}  
A plot of the $m_{16}\ vs.\ m_{1/2}$ plane, illustrating  
contours of Yukawa unification parameter R(black),  
 $BF(b\to s\gamma )(\times 10^4)$ (magenta), $a_\mu(\times 10^{10})$ (blue),   
$BF(B_s\to \mu^+\mu^-)(\times 10^{8})$ (light blue) and $\Omega_{\tz_1}h^2$ (green).}}   
  
Phenomenologically,   
the most problematic is the value of the neutralino relic density  
$\Omega_{\tz_1}h^2$. For pt. 1, $\Omega_{\tz_1}h^2$ is in the  
observationally  acceptable range $\Omega_{CDM} = 0.1-0.3$ because  
of resonance annihilation via $\tz_1\tz_1\to h\to b\bar{b}$ in the early  
universe.\footnote{Although $2m_{\tz_1}$ is several widths below $m_h$,  
thermal motion can lead to resonance enhancement.} 
Annihilation via $t$ and $u$ channel sfermion exchange is also significant.   
As $m_{16}$ increases (pts. 2-4), these other annihilation channels are   
increasingly suppressed. Furthermore, for these points,   
$2m_{\tz_1} > m_h$ so that  annihilation via $s$-channel $h$ is no  
longer resonant. As a result, the neutralino relic   
density is very high. There is, therefore, considerable tension between the  
theoretical requirement of a high degree of Yukawa coupling unification, and  
the phenomenological requirement  
that the age of the universe is greater than 10 Gyr.
There seems to be little overlap of  
parameter regions consistent with both requirements  
for these $\mu >0$ solutions.\footnote{To obtain a relic density in an  
acceptable range by resonant annihilation via $h$, the parameters have  
to fall in a narrow range. This is clear for $m_{1/2}$ which controls  
the LSP mass, but we found that even changing $m_{10}$, $M_D$ and $A_0$ to   
$1.24m_{16}$, $0.32m_{16}$ and $-2m_{16}$, respectively  
(compare these with corresponding values  
in Fig.~\ref{HSplane_2}) $\Omega h^2$ in the $h$ corridor is never less than   
0.3, and is typically closer to unity.}  
(In the case of $\mu <0$, a reasonable  
relic density can be obtained by large rates for   
$\tz_1\tz_1\to A,\ H\to b\bar{b}$ in the early  
universe\cite{Baer:2000jj}, but then the problem is with the $b \to  
s\gamma$ decay rate.)  
  
It is well known that the relic density can be low in the ``focus point''  
region of the mSUGRA model\cite{focus}. The focus point region occurs near the   
upper boundary of the common scalar mass $m_0$, which is determined by  
where $\mu^2\to $ negative values, thus signaling a breakdown in REWSB.  
Since $|\mu |$ is small, the $\tz_1$ has a significant higgsino component,  
and there is large annihilation to states such as $WW$ or $ZZ$, and  
possibly even co-annihilation if $m_{\tw_1}\simeq m_{\tz_1}$.    
A possible solution to the too-high relic density problem is to try to  
dial in lower Higgs mass splittings. This should make the model more   
mSUGRA-like, and bring back the focus-point region.   
We have checked this,   
but found that increasing the higgsino component of the lighter  
charginos and neutralinos also modifies the $b$ and $t$ quark  
threshold corrections so that Yukawa unification can reach only values of  
$R=1.2-1.3$.  
  
The other option to lower the relic density is to find a region with  
neutralino annihilation via the light Higgs boson. 
Such an example is presented by point~1 of Table~\ref{tab:one} and  
illustrated in Fig.~\ref{HSplane_2}. One can clearly see a
corridor in $m_{1/2}$ of $\sim 15$ GeV width where 
annihilation through the the light Higgs  
boson occurs. Part of this corridor  
satisfies all experimental constraints but Yukawa unification reaches  
only 19\% here.

\section{Supersymmetric models with $\mu <0$}  
  
Next, we turn to the re-examination of Yukawa unification  
for models with $\mu <0$. Our goal is to update our earlier  
results\cite{Baer:1999mc,Baer:2000jj}, which found  
Yukawa unification to be possible for a range of parameters  
within the $DT$ model.

\subsection{mSUGRA model}  
  
Again, we begin by examining Yukawa unification within the mSUGRA  
model. As before, we scan over the four mSUGRA model   
parameters, and plot the value of $R$ against each of them.  
Our results are shown in Fig.~\ref{fig:sugm}.   
\FIGURE[t]{  
\epsfysize=14cm\epsfbox[0 25 567 540]{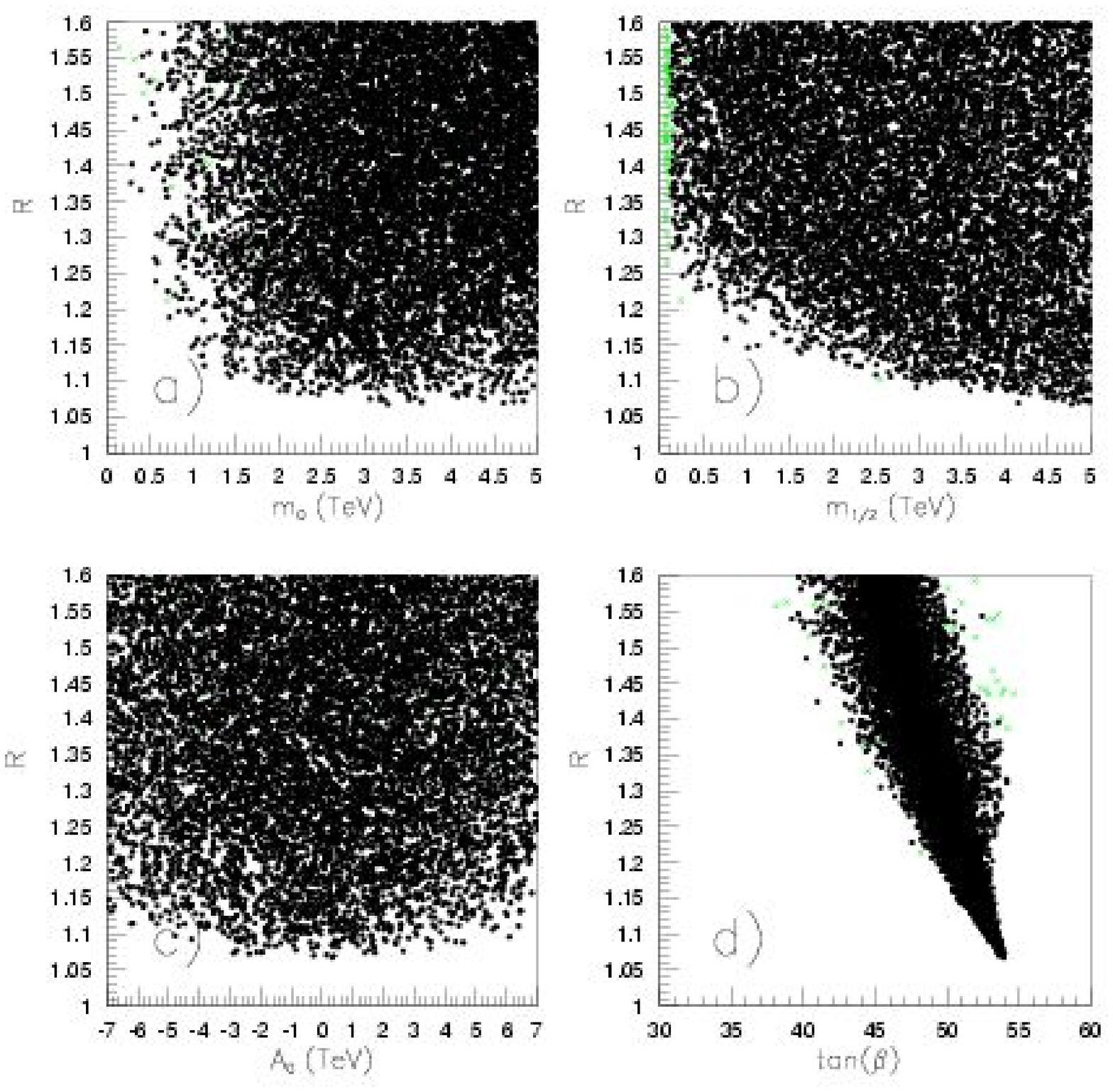}  
\caption{Plot of Yukawa unification parameter $R$ versus  
input parameters of the mSUGRA model for {\it a}) $m_0$,   
{\it b}) $m_{1/2}$, {\it c}) $A_0$ and {\it d}) $\tan\beta$,  
when $\mu <0$.}  
\label{fig:sugm}}  
The    
results of these scans show that Yukawa unification to $R\sim 1.15-1.20$  
can be achieved if $m_0,\ m_{1/2} \alt 1$ TeV. But as model  
parameters increase, the Yukawa unification improves, and is  
reaching the 5--10\% level for $m_{1/2}\sim 5$ TeV  
and $\tan\beta \sim 55$. To see whether unification is possible for yet  
larger values of sparticle masses,    
we continued these scans out to even higher values of model  
parameters. We found that Yukawa unification   
{\it is possible} within the mSUGRA model for $m_0$ values  
around 10 TeV, and $m_{1/2} \sim 1.6 m_0$.   
Our results for the extended mSUGRA parameter space  
illustrating this are shown in Fig.~\ref{fig:sugmex}.  
\FIGURE[t]{  
\epsfysize=13cm\epsfbox[80 20 510 600]{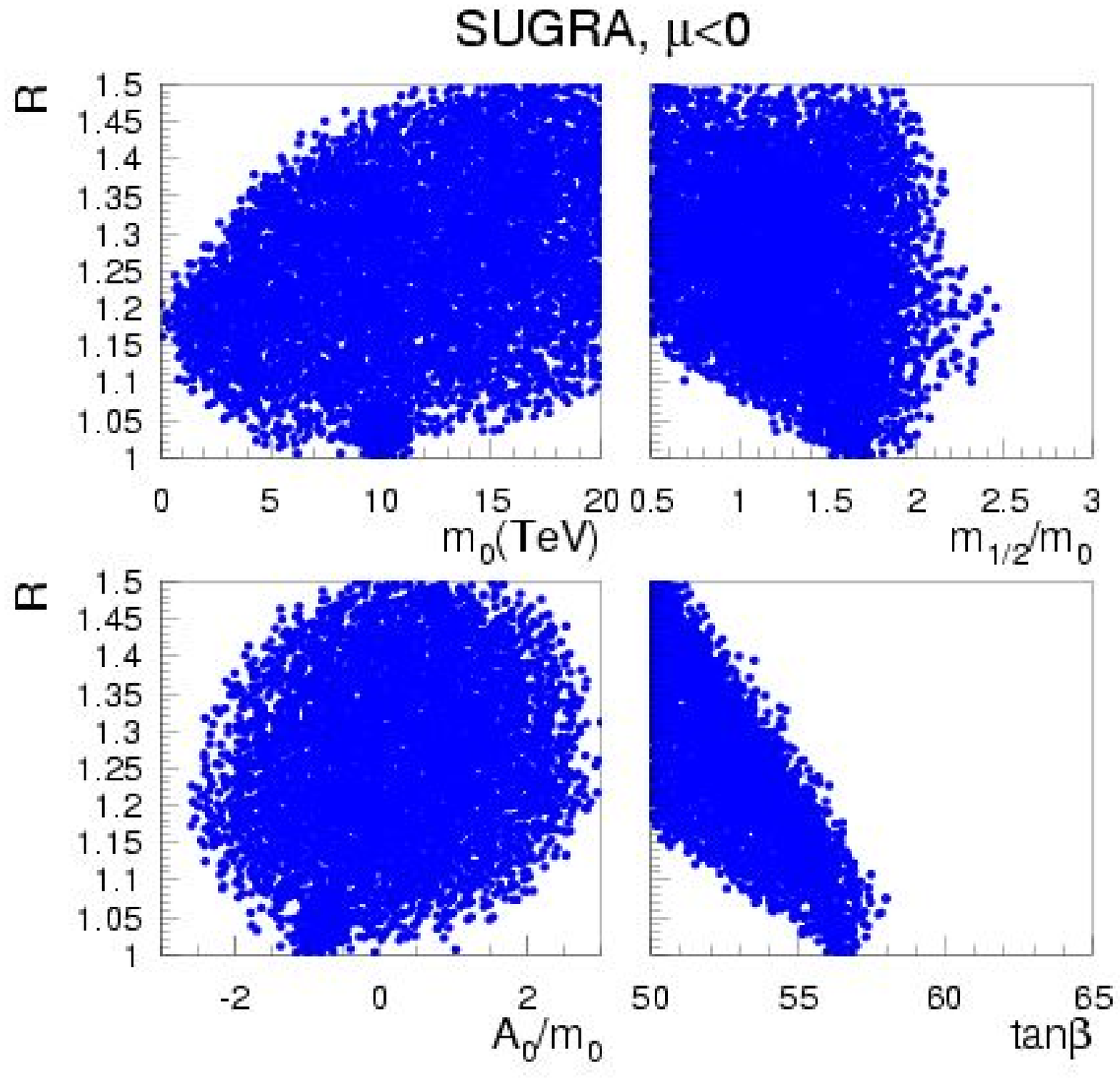}  
\caption{Plot of Yukawa unification parameter $R$ versus  
input parameters of the mSUGRA model,    
{\it a}) $m_0$,   
{\it b}) $m_{1/2}$, {\it c}) $A_0/m_0$ and {\it d}) $\tan\beta$,  
when $\mu <0$ and extended ranges of these parameters.}  
\label{fig:sugmex}}  
A particular case is shown as point 1 in  
Table \ref{tab:two}, which has Yukawa unification to 1\%.   
In this case, squarks and sleptons have mass  
values of 10--30~TeV (including third generation sparticles), and   
the models suffer from the need for fine-tuning.   
The models are in the  
decoupling regime, so that the values  
of $a_\mu$ and $BF(b\to s\gamma )$ will be near the SM predicted  
values. However, the neutralino relic density is extremely high  
($\Omega h^2 =52.6$ for the model listed in the table), so that  
$R$-parity violation would be needed to make the model  
cosmologically viable. Of course,    
we would then need another candidate   
for dark matter.

\begin{table}  
\begin{center}  
\caption{\label{tab:two}Model parameters, Yukawa couplings and weak scale   
sparticle masses for four case studies for $\mu <0$.}  
\bigskip  
\begin{tabular}{lccccc}  
\hline  
                                 & 1	  & 2	   & 3         & 4       & 5       \\  
parameter                        & mSUGRA & $DT_1$ &$DT(f_\nu)$& $DT_2$  & $HS$  \\  
\hline                                                                            
$m_{16}$                         &  9616.1& 1000.0 &    1000.0 & 2775.8 & 2000.0 \\  
$m_{10}$                         &  9616.1& 1414.0 &    1414.0 & 3763.6 & 2709.4 \\  
$M_D   $                         &  0.0   &  333.0 &     333.0 & 1095.0 & 1048.4 \\  
$m_{1/2}$                        & 14893.2&  700.0 &     700.0 & 2033.6 & 1500.0 \\  
$A_0$                            & -7943.8&    0.0 &       0.0 &    0.0 & -560.2 \\  
$\tan\beta$                      &  56.29 &   54.0 &      54.0 &   56.0 &   54.8 \\  
$M_N$                           &$M_{GUT}$&$M_{GUT}$&                         
$1.0\times 10^{15}$ & $M_{GUT}$  &$M_{GUT}$   \\                                   
$f_t(M_{GUT})$                   &  0.610 &  0.599 &     0.609 &  0.611 &  0.608 \\  
$f_b(M_{GUT})$                   &  0.606 &  0.556 &     0.580 &  0.614 &  0.616 \\  
$f_\tau (M_{GUT})$               &  0.609 &  0.572 &     0.582 &  0.616 &  0.609 \\  
$R$                              &   1.01 &   1.08 &      1.05 &   1.01 &   1.01 \\  
$m_{\tg}$                        & 27536.5& 1602.2 &    1600.9 & 4319.5 & 3241.9 \\  
$m_{\tu_L}$                      & 24898.2& 1697.2 &    1696.5 & 4641.9 & 3355.9 \\  
$m_{\td_R}$                      & 23573.9& 1518.1 &    1517.8 & 3960.2 & 3237.2 \\  
$m_{\tst_1}$                     & 19073.3& 1166.2 &    1169.1 & 3304.4 & 2388.5 \\  
$m_{\tb_1}$                      & 19381.7& 1083.6 &    1069.1 & 2704.3 & 2328.8 \\  
$m_{\tell_L}$                    & 13292.6&  928.5 &     928.5 & 2376.3 & 2192.2 \\  
$m_{\tell_R}$                    & 10981.9& 1094.4 &    1094.4 & 3108.4 & 2121.3 \\  
$m_{\tnu_{e}}$                   & 13292.4&  925.1 &     925.0 & 2375.0 & 2190.7 \\  
$m_{\ttau_1}$                    &  7843.1&  715.2 &     694.6 & 1795.7 & 1268.6 \\  
$m_{\tnu_{\tau}}$                & 12166.8&  734.6 &     704.9 & 1796.8 & 1839.3 \\  
$m_{\tw_1}$                      & 12415.4&  419.4 &     448.1 & 1344.7 & 1179.8 \\  
$m_{\tz_2}$                      & 12415.3&  422.3 &     449.9 & 1345.5 & 1309.9 \\  
$m_{\tz_1}$                      & 7143.9 &  288.4 &     290.0 &  894.5 &  654.0 \\  
$m_h      $                      &  127.6 &  120.0 &     120.0 &  123.9 &  123.5 \\  
$m_A      $                      &  1854.8&  624.3 &     614.9 & 1864.4 & 1384.0 \\  
$m_{H^+}  $                      &  1859.2&  632.9 &     623.6 & 1868.7 & 1388.9 \\  
$\mu      $                      &-12943.7&  -438.1&    -472.7 &-1353.7 &-1277.5 \\  
$S        $                      &   1.46 &  1.80  &     1.81  &  1.80  &  1.79  \\  
$a_\mu \times 10^{10}$           & -0.0453&-40.8   &   -40.0   &-10.3   & -2.76  \\  
$BF(b\to s\gamma)\times 10^4$    &   3.52 &  7.35  &     7.40  &  4.08  &  4.47  \\  
$BF(B_s\to\mu^+\mu^-)\times 10^8$&   0.198&  0.273 &     0.300 &  0.315 &  0.223 \\  
$\Omega_{\tz_1}h^2$              &  52.6  & 0.00854&    0.00683&  0.251 &  0.294 \\  
\hline  
\end{tabular}  
\end{center}  
\end{table}  
  
\subsection{$DT$ model}  
  
Next, we re-examine Yukawa coupling unification in the $DT$ model with  
$\mu <0$. We scan over similar ranges of model parameters as in the $\mu  
>0$ case, and plot the value of $R$ versus model parameters in  
Fig. \ref{fig:DTmundots}.  As for the mSUGRA model just discussed,    
we find that a high  
degree of Yukawa coupling unification is possible for $\tan\beta\sim  
55$.  However, in contrast to the mSUGRA model case, we can now achieve  
Yukawa coupling unification even for $m_{16}$ values of 1 TeV, or   
smaller. Yukawa unified models prefer large values of $m_{1/2}$,  
typically greater than a TeV.  Unlike the Yukawa unified models with  
$\mu >0$, a wide range of ratios of $A_0/m_{16}$ are allowed, centered  
about $A_0\sim 0$. Moreover, the parameter $m_{10}$ can be greater  
than or less than $m_{16}$. The final frame of Fig. \ref{fig:DTmundots}  
shows that a positive $D$-term $M_D/m_{16} \sim 0.2-0.4$
is required for a  
high degree of Yukawa coupling unification.    
The need for the $D$-term is lessened as we increase the range of model  
parameters, the limiting case being the previously shown mSUGRA model.  
  
\FIGURE[t]{\hspace*{-.5cm}\epsfig{file=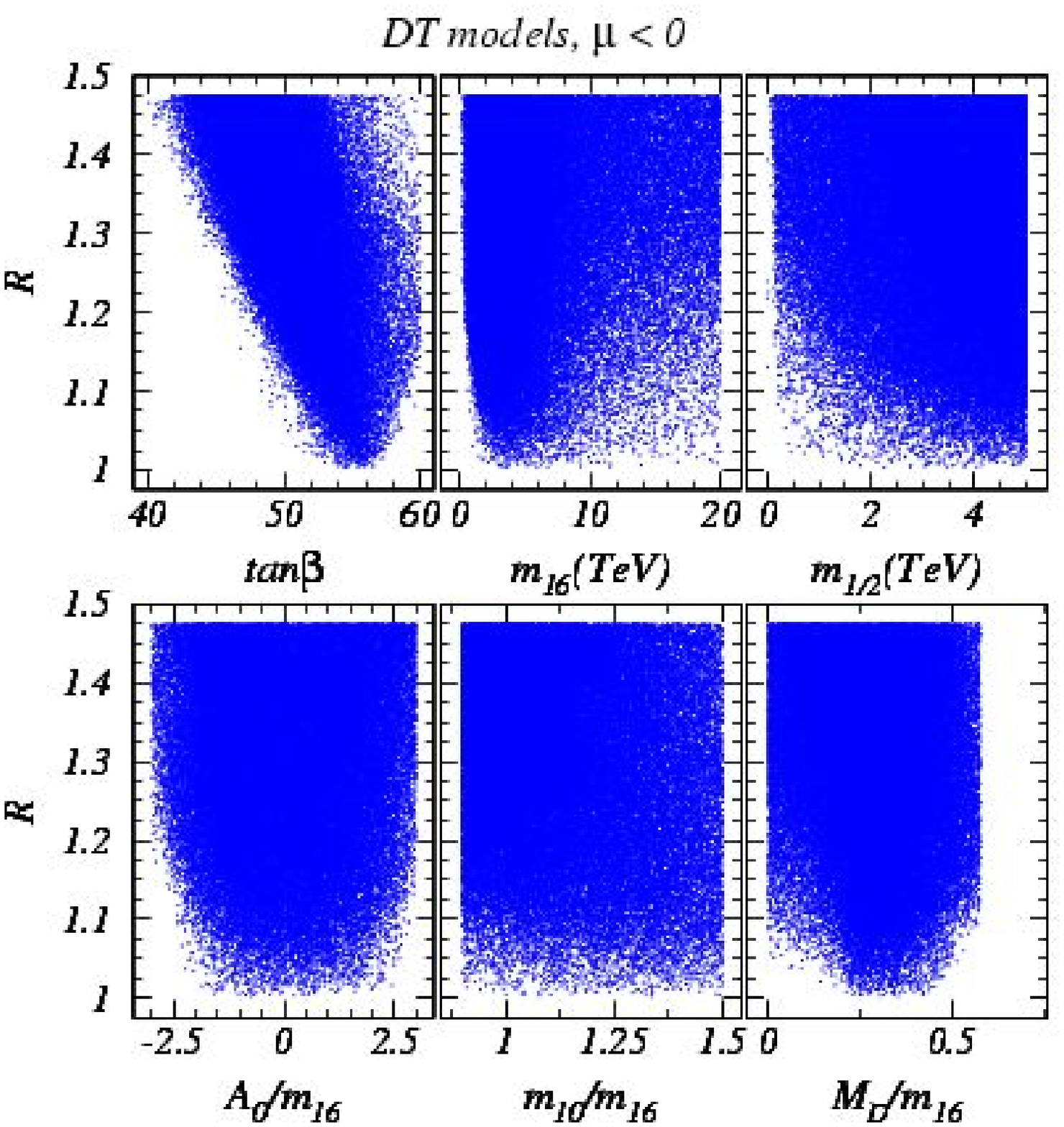,height=15cm}  
\caption{  
Plot of $R$ versus parameters of the $DT$ model with $\mu <0$   
for  
{\it a}) $\tan\beta$,   
{\it b}) $m_{16}$, {\it c}) $m_{1/2}$, {\it d}) $A_0/m_{16}$,   
{\it e}) $m_{10}/m_{16}$ and {\it f}) ${\rm sign}(M_D^2)\sqrt{|M_D^2|}/m_{16}$.}  
\label{fig:DTmundots}}

In Fig. \ref{fig:fulld}, we show contours for the same   
observables as in Fig.~\ref{HSplane_2} along  
with contours of $R$ in the $m_{16}\ vs.\ m_{1/2}$  
plane for $\tan\beta$ values of $52,\ 54,\ 56$ and $58$.  
The legends for the various contours are shown on the figure.  
We take $m_{10}=1.414 m_{16}$, $M_D=0.333 m_{16}$ and $A_0=0$.   
In the figures, the low $m_{16}$ regions of parameter space  
are excluded by the  requirement $m_{\ttau_1}>m_{\tz_1}$. As $\tan\beta$  
increases, this excluded region   
is usurped by the requirement  
that $m_A^2>0$, which increasingly excludes the parameter plane denoted  
by asterisks.  
In the large $m_{16}$ and low $m_{1/2}$ regions,   
parameter space is forbidden because $\mu^2 <0$ (pluses), which signals  
a breakdown in REWSB. Thus, the allowed parameter space  
turns out to be a narrow wedge between these two extremes, and  
is increasingly restricted as $\tan\beta$ increases.  
  
\FIGURE[t]{  
\epsfig{file=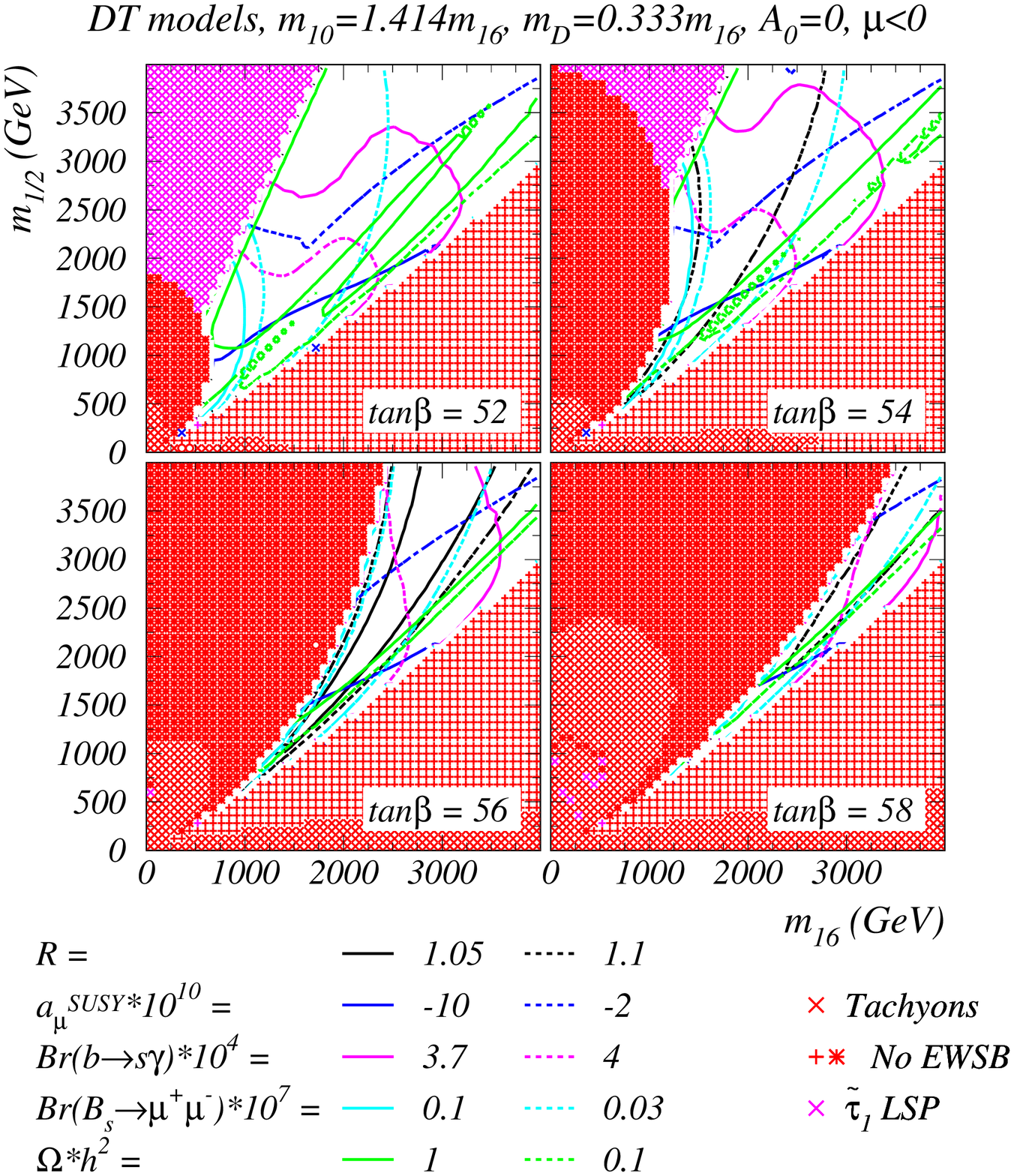,height=17cm}  
\caption{Contours of $R$ together with those of  
selected observables in $m_{16}\ vs.\ m_{1/2}$ plane for various   
$\tan\beta$ values, with $\mu <0$ and $A_0=0$ for the $DT$ model.}  
\label{fig:fulld}}  
  
In the theoretically allowed regions, the solid and dotted black contours show  
regions where $R< 1.05$ and 1.10, respectively. The largest  
Yukawa unified regions occur in the third frame, where $\tan\beta =56$.  
In between the solid (dashed) lines for this case, $R < 1.05$ (1.1). In  
the neighbouring cases with $\tan\beta=54$ and $\tan\beta = 58$, no  
solid black line appears because $R> 1.05$, while in the first frame  
$R > 1.1$.  
We also show contours of values of $a_\mu$, $BF(b\to s \gamma )$,  
$BF(B_S\to \mu^+\mu^- )$ and $\Omega_{\tz_1}h^2$. The   
$a_\mu$ values are always negative, and so disfavored by   
recent results from the E821 experiment. However, for  large  
values of model parameters, $a_\mu$ approaches its SM value, and  
lies at least within $3\sigma$ of the experimental result  
(the exact deviation depends upon which SM $a_\mu$ calculation  
is adopted). Even more problematic is the value of   
$BF(b\to s\gamma )$, which is above $4\times 10^{-4}$ below  
the dashed magenta contours. Requiring consistency of the   
models with $BF(b\to s\gamma )$ pushes model parameters to very high  
values. Contours of $BF(B_s\to\mu^+\mu^- )=0.1\times 10^{-7}$ and  
$0.03\times 10^{-7}$ are also shown; this branching fraction seems somewhat  
out of reach of the CDF experiment.   
  
The values $\Omega_{\tz_1}h^2 =0.1$ and 1 are also shown by the green  
dashed and solid contours. In the frames shown, a reasonable  
relic density is obtained in three distinct regions.  
\begin{enumerate}  
\item Near the boundary of the low $m_{16}$ excluded region (shown by  
magenta x's)  
for $\tan\beta =52$ and 54, $m_{\ttau_1}\simeq m_{\tz_1}$ and  
$\ttau_1 -\tz_1$ and $\ttau_1 -\ttau_1$ co-annihilation  
reduces the relic density to reasonable values\cite{ellis}.  
\item Near the boundary of large $m_{16}$, where $|\mu |$ becomes  
small, the higgsino component of $\tz_1$ increases, so that there is  
efficient $\tz_1 -\tz_1$ annihilation into vector boson pairs.  
\item In the first three frames, a corridor of resonance annihilation  
$\tz_1\tz_1\to A,\ H\to b\bar{b},\ \tau\bar{\tau}$ occurs, which  
severely lowers the relic density\cite{dn}.  
\end{enumerate}  
  
Simultaneously fulfilling all three constraints, $a_\mu$,   
$BF(b\to s\gamma )$ and $\Omega h^2$, plus obtaining a high degree  
of Yukawa coupling unification is challenging--but possible--and  
depends upon the tolerances required of each constraint.  
  
As an example, we show point 2 in Table \ref{tab:two}, with good Yukawa   
unification and relatively low values of soft SUSY breaking model
parameters. For this point, third generation squark masses are in the   
1-2 TeV range, so   
there should not be too much fine tuning required.   
The sparticle mass spectrum is  
rather heavy, and other than possibly the light boson Higgs $h$, none of  
the new states would be  
accessible to Tevatron SUSY searches. Sparticles   
should be accessible to LHC SUSY searches, but would be inaccessible   
at 500~GeV linear colliders (except for the Higgs $h$). As the machine energy
is increased, $\tz_1\tz_2$ production 
would be accessible for $\sqrt{s} \agt 750$~GeV, and 
chargino pair production would become accessible for $\sqrt{s}\agt 850$ GeV.
The value of $m_{\tz_1}\simeq m_A/2$ (given the broad width of the  
$A$ and $H$), so that efficient $s$-channel annihilation of neutralinos  
can take place in the early universe, and a very low relic density   
is generated. However, the values of $a_\mu$ and $BF(b\to s\gamma )$  
are well beyond their experimental limits, so that the point is   
likely excluded.  
  
The third point in Table \ref{tab:two} shows the same parameter  
space point, except that the MSSM is enlarged to include a gauge   
singlet chiral scalar superfield $\hat{N}_i^c$ for each generation  
$i=1-3$, and which   
contains a right-handed neutrino  
as its fermionic component. The superpotential is enlarged to include  
the terms  
\begin{equation}  
\hat{f}\ni {1\over 2} M_N \hat{N}^c\hat{N}^c+  
f_\nu  \epsilon_{ij}\hat{L}^i\hat{H}_u^j\hat{N}^c  
\end{equation}  
for each generation. We retain only the third generation Yukawa coupling  
which we assume satisfies $f_{\nu_{\tau}} =f_t$ at  
$Q=M_{GUT}\simeq 2\times 10^{16}$ GeV, and take   
$M_N =1\times 10^{15}$ GeV. 
This value of $M_N$ gives  
a $\nu_\tau$ mass in accord with atmospheric neutrino  
measurements at the SuperK experiment. The neutrino Yukawa coupling  
is coupled through the RGEs with the other Yukawa couplings and  
soft SUSY breaking parameters\footnote{We use two loop  
RGE evolution of the MSSM+RHN model as given in the  
second paper of Ref. \cite{imh}.},  
but decouples below the scale $M_N$. Its effect, in this case,  
is to {\it improve} Yukawa coupling unification from $R=1.08$ to  
$R=1.05$. In addition, the third generation neutrino Yukawa coupling acts to  
somewhat suppress the third generation slepton and sneutrino masses.   
  
We found that it is possible to reasonably satisfy the indirect  
experimental constraints while obtaining good Yukawa unification in the  
$DT$ model provided that the input parameters with mass dimensions are  
increased. This is illustrated by point 4 of Table \ref{tab:two}. Since  
$m_{16}$ and $m_{1/2}$ (and proportionally $m_{10}$ and $M_{D}$) are  
somewhat higher than in point 2, the mass spectrum is heavier and only  
the lightest Higgs particle is detectable at the Tevatron or the LC. But  
the Yukawa unification is perfect and the neutralino relic density is  
ideal. The values of $a_\mu$ and $BF(b\to s\gamma)$ comply substantially  
better with the experiments than those of point 2 or 3, due to the  
heaviness of superpartners in the relevant loops. Unfortunately,  
direct detection of sparticles will be difficult even at the LHC.   
  
\subsection{$HS$ model}  
  
We also investigate Yukawa coupling unification for $\mu <0$  
in the $HS$ model. We scan over parameter space exactly as in the  
$DT$ model, except that mass splittings are only applied to the Higgs  
multiplets. The results of $R$ versus various model parameters  
are shown in Fig. \ref{fig:HSmundots}.  
We find that Yukawa coupling unification is again possible below the  
1\% level for a wide range of model parameters. In fact, the results  
are qualitatively quite similar to the case of the $DT$ model for $\mu <0$.  
The best unification occurs at $\tan\beta\sim 55$. The main difference is   
that a wider range of $D$-terms is now allowed.  This, it has been  
pointed out\cite{bdr}, is because in the $HS$ scenario   
sfermion masses and hence the corrections to $m_b$ are not altered by  
the splitting: in contrast, in the $DT$ scenario, $m_{\tb_R}^2$ is reduced  
by positive $D$-terms, resulting in an enhancement of the gluino  
contribution to $\delta m_b$.  
  
\FIGURE[t]{\hspace*{-.5cm}\epsfig{file=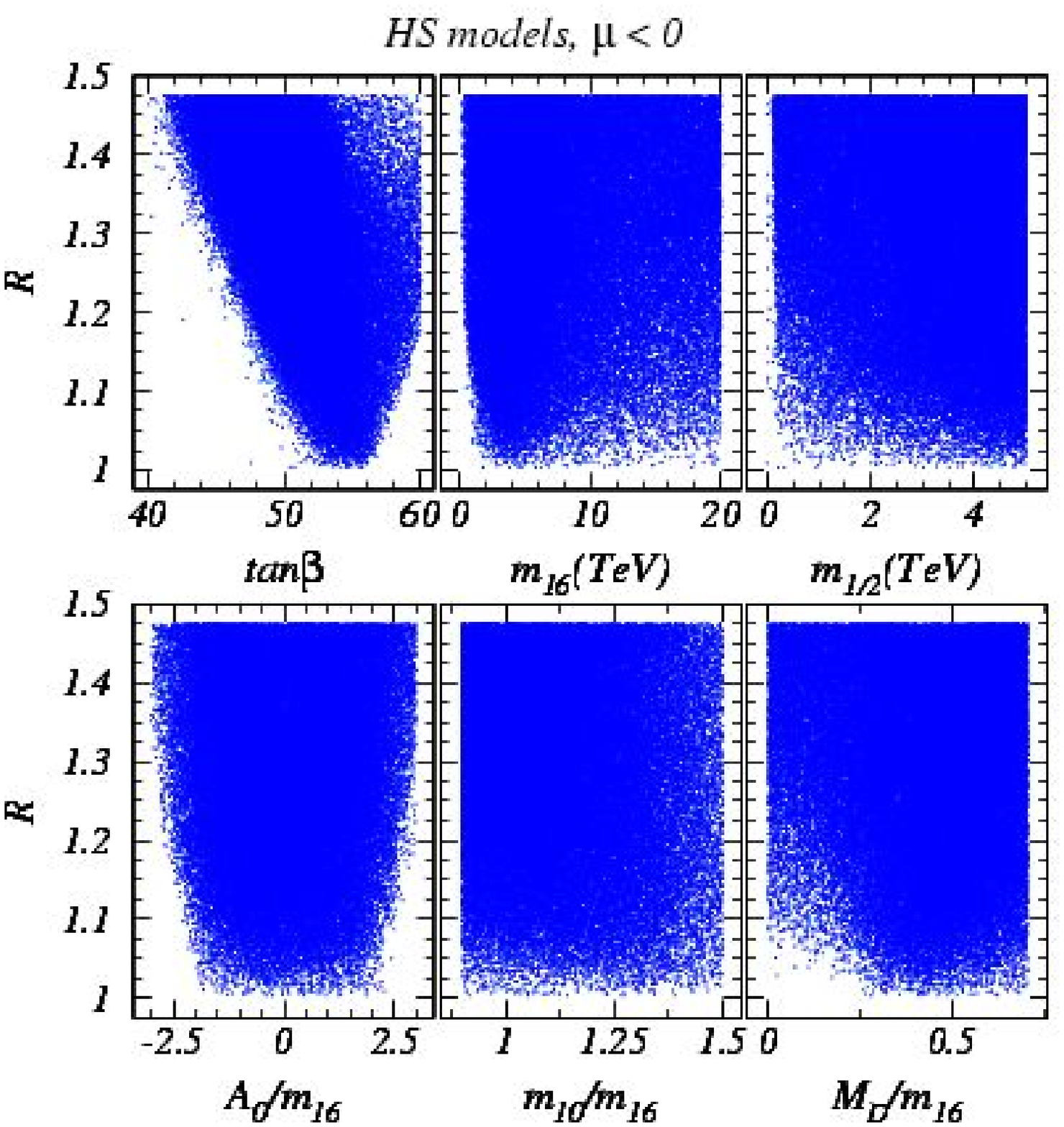,width=16cm}  
\caption{ The unification parameter $R$ versus parameters    
{\it a}) $\tan\beta$, {\it b}) $m_{16}$, {\it  
c}) $m_{1/2}$, {\it d}) $A_0/m_{16}$, {\it e}) $m_{10}/m_{16}$ and {\it  
f}) ${\rm sign}(M_D^2)\sqrt{|M_D^2|}/m_{16}$ of the $DT$  
model, for $\mu <0$.}  
\label{fig:HSmundots}}  
  
In Fig. \ref{fig:HSmun}, we show the same $m_{16}\ vs.\ m_{1/2}$ plane  
plots as in Fig.~\ref{fig:fulld}, but for the $HS$ model.    
Again, we show contours of Yukawa coupling  
unification parameter $R$,   
$a_\mu$, $BF(b\to s\gamma )$, $BF(B_s\to \mu^+\mu^- )$ and  
$\Omega_{\tz_1}h^2$.  The results are qualitatively very similar to  
those of the $DT$ model, except that the Yukawa coupling unification is  
marginally worse, and the allowed range of parameters is slightly reduced.

%
%
We show point~5 in Table \ref{tab:two} as an example of an $HS$ model. This   
model point also features more massive sparticles than points 2 and 3, and is   
inaccessible at the Tevatron, and probably, even at the LHC.  
This point has essentially perfect Yukawa unification and   
satisfies all the experimental data within experimental and theoretical  
errors.  
  
\FIGURE[t]  
{\hspace*{-1cm}\epsfig{file=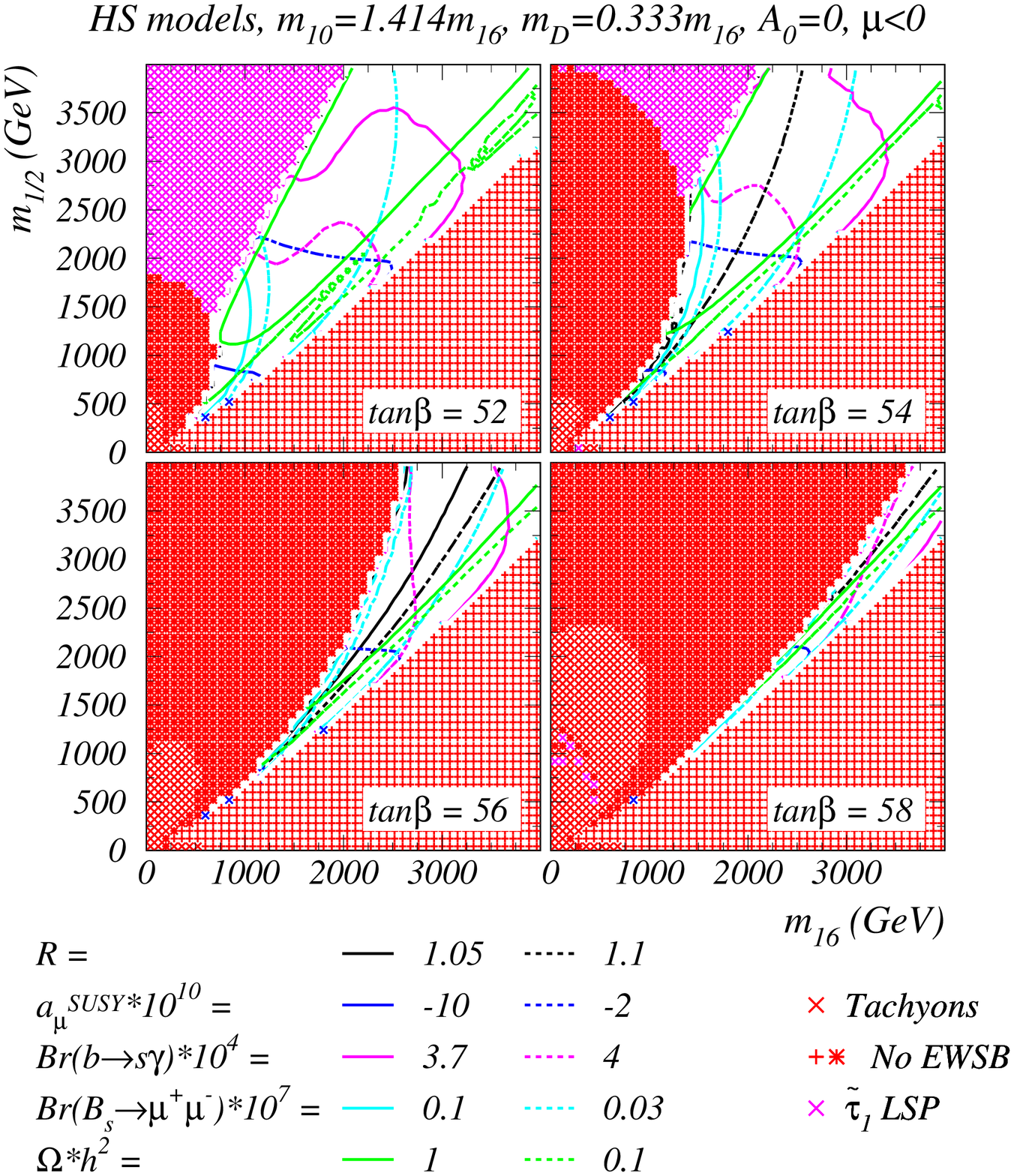,width=16cm}  
\caption{Plot of $m_{16}\ vs.\ m_{1/2}$ plane for various   
$\tan\beta$ values, $\mu <0$ and $A_0=0$.}  
\label{fig:HSmun}}  
  
\section{Comparison with BDR results}\label{compare}  
  
As we have already discussed, a similar analysis to the one  
in this paper has been performed by  
Blazek, Dermisek and Raby (BDR)\cite{bdr} for the case  
of $\mu >0$. The results of BDR have  
many similarities, but also important differences, to our results   
and to the results in Ref.~\cite{Baer:2001yy}.   
In this section, we  first compare our calculational   
procedure based on ISAJET with that of BDR. Next, we  
comment on similarities and differences in numerical results. Finally,   
in the last section, we  
present some general remarks about our    
program of calculations.

\subsection{Comparison of the top--down versus bottom--up procedures}  
  
BDR adopt a top-down approach to calculating  
the sparticles mass spectrum. They begin with the parameters,  
\begin{equation}  
M_{GUT},\ \alpha_{GUT},\ \epsilon_3,\ \lambda,\ \mu,\ m_{1/2},\ A_0,\  
\tan\beta,\ m_{16}^2,\ m_{10}^2,\ \Delta m_H^2,  
\end{equation}  
where $\epsilon_3$ parameterizes the small $GUT$ scale non-unification
of the $SU(3)_C$ gauge coupling with the electroweak gauge couplings
($\epsilon_3=(\alpha_3(M_{GUT})-\alpha_{GUT})/\alpha_{GUT}$), $\lambda $
is the unified third generation Yukawa coupling, $\Delta m_H^2$
represents the Higgs mass squared splitting (about $m_{10}^2$) in the
$HS$ model (equal to $M_D^2$ in the $DT$ model), and the other
parameters are as in this paper. They evolve via two-loop RGEs for
dimensionless parameters and one-loop RGEs for dimensionful parameters
via MSSM RGEs to the scale $Q=M_Z$, where REWSB is imposed assuming the
radiatively corrected scalar potential of a two-Higgs doublet model with
softly broken supersymmetry.
At $Q=M_Z$, they evaluate the 9 observables: $\alpha_{EM}$, $G_\mu$,   
$\alpha_s(M_Z^2)$, $M_Z$, $M_W$, $\rho_{NEW}$, $m_t$,   
$m_b^{\overline{MS}}(m_b)$ and $m_\tau$, all including   
one-loop radiative corrections.   
In particular, the SUSY threshold corrections to $m_t$, $m_b$ and $m_\tau$  
are all calculated and imposed at scale $Q=M_Z$.  
Also, the spectrum of SUSY and   
Higgs particles is calculated.   
Radiative corrections are included for the various Higgs   
masses $m_h$, $m_H$ and $m_A$, while SUSY particle masses are   
taken to be running mass parameters evaluated at $Q=M_Z$. 
Next, the experimental values for these  
observables together with the associated errors are used to  
obtain  $\chi^2$ for each set of inputs,  
and this $\chi^2$ is minimized using the CERN program   
MINUIT. Thus, by scanning over the 11-dimensional GUT scale parameter  
space, they search for parameter regions leading to good agreement  
with the 9 observables. In particular, their third generation   
Yukawa couplings are  
always truly unified. In general, the electroweak observables and the   
third generation fermion masses  
will deviate from their measured central values.  
BDR seek input parameter choices which minimize the overall   
deviation, and result in acceptable values of $\chi^2$.   
  
We use ISAJET, where an iterative, bottom-up approach is adopted. The  
calculation begins by inputting at scale $Q=M_Z$ the $\overline{DR}$  
central values of the three gauge couplings $g_1$, $g_2$, $g_3$, and the  
two-loop $\overline{DR}$ central values of fermion masses $m_b$ and  
$m_\tau$. The two-loop value of $m_t^{\overline{DR}}$ is input at scale  
$Q=m_t$.  In the first iteration, the gauge and Yukawa couplings are  
evolved via 2-loop RGEs to the scale $Q=M_{GUT}$\footnote{$M_{GUT}$ is
defined as the scale at which $g_1=g_2$. The strong coupling $g_3$
remains un-unified, so for ISAJET, the BDR parameter $\epsilon_3$
is effectively an output, and not an adjustable parameter.}, 
beginning first with  
SM RGEs, then transitioning to MSSM RGEs at a scale $M_{SUSY}$, typical  
of the expected sparticle masses.  At $M_{GUT}$, the various soft SUSY  
breaking masses are included, and the set of 26 2-loop MSSM RGEs are  
used for evolution back to the weak scale. The various soft SUSY  
breaking masses are frozen out at scales $Q$ equal to their mass to  
minimize logarithmic radiative corrections. The 1-loop RG improved  
effective potential is minimized at the scale  
$Q=\sqrt{m_{\tst_L}m_{\tst_R}}$, where the requirement of  
REWSB is imposed, and the  
value of $\mu^2$ (along with bilinear soft term $B$) is {\it derived}.  
  

On this and subsequent iterations, the 1-loop logarithmic SUSY threshold  
corrections to gauge and Yukawa couplings 
are included through RGE decoupling, {\it i.e.}   
by changing the corresponding $\beta$  
functions as various sparticle thresholds are passed, making a smooth  
transition from the MSSM to the SM, or vice-versa.  Then, the 1-loop  
finite SUSY threshold corrections to $m_b(M_Z)$, $m_\tau(M_Z)$ and $m_t(m_t)$  
are computed, and the corresponding Yukawa couplings are determined. Finally,
all Yukawa couplings and SUSY breaking parameters required to evaluate the  
threshold correction to fermion masses are extracted at the scale  
$M_{SUSY}=\sqrt{m_{\tst_L}m_{\tst_R}}$, or at the scale associated with  
each loop contribution ({\it e.g} the gluino loop contribution  
to $\delta m_b$ is evaluated using  
values of $\alpha_s$ and $M_3$ evaluated at the scale $Q=m_{\tg}$).

Radiative corrections are included in calculating the Higgs masses  
and couplings and $m_{\tg}$, but not for other sparticle   
masses.  
Using updated Yukawa couplings, RGE running between $M_Z$ and $M_{GUT}$
and back is iterated until a convergent solution within tolerances is
achieved.  In this approach, the third generation fermion masses, $M_Z$
and $\sin^2\theta_W$ are fixed at their central value, but the Yukawa
couplings do not unify perfectly. Regions of parameter space which lead
to Yukawa coupling unification within specified tolerances are then
searched for.

%
%
   
An important ingredient for accurately obtaining the   
superpotential Yukawa couplings is  
the evaluation of SUSY threshold corrections to the SM fermion masses. In  
Fig. \ref{fig:loop}, we show an  
illustrative example of these self energy corrections for the top  
and bottom quarks, denoted by $\Sigma_t$ and $\Sigma_b$, respectively,  
for both the $DT$ model  
(dashes) and $HS$ model (solid), versus the parameter $m_{16}$.   
Other parameters are specified on the figure.  
Here,  
the $\Sigma$s are defined through their relation with the pole fermion  
masses,  
\begin{equation}  
m_f^{pole}=m_f^{\overline{DR}}(Q) \left( 1 +\Sigma_f (Q) \right) .  
\end{equation}  
where $m_f^{\overline{DR}}(Q)$ would be  
computed using the running Yukawa coupling and the vacuum expectation value   
at the relevant scale. For instance for the top quark mass  
$m_t^ {\overline{DR}}(m_t) = f_t(m_t) v(m_t) \sin \beta$.

  
In frames {\it a}) and {\it b}), we show contributions to $\Sigma_t$ from  
$\tg\tst_i$ loops (red) and $\tw_i-\tb_j$ and other loops (green) (the sum of
the green curves is shown by the blue curve)
and the total contribution (black). 
Frame {\it a}) is for  $\mu >0$,  
while for frame {\it b}) $\mu <0$.  
From the first two frames, we see that the gluino--stop loops provide the  
dominant contribution to the top quark self energy, but that in moving   
from the $DT$ to the $HS$ model, there is relatively little change.  
Moreover, except for the smallest values of $m_{16}$ and positive $\mu$,  
the total SUSY contribution to $\Sigma_t$ is quite stable around $(4\pm 1)$\%.

  
In frames {\it c}) and {\it d}), we show the corresponding SUSY
contributions to the $b$-quark self energy. The finite corrections
coming from gluino--sbottom, $\Sigma_b (\tg\tb_j)$, and chargino--stop
loops, $\Sigma_b (\tw_i\tst_j)$, provide the dominant contributions, and
are approximated at large $\tan\beta$ by,
\begin{eqnarray}
\Sigma_b (\tg\tb_j) &\simeq& \frac{2}{3} \frac{\alpha_s (m_{\tg})}{\pi}
m_{\widetilde{g}} \mu \tan\beta\ C( m^2_{\widetilde{b}_1},
m^2_{\widetilde{b}_2}, m^2_{\widetilde{g}} ), \nonumber \\
\Sigma_b(\tw_i\tst_j) &\simeq& \frac{f_t^2}{16
\pi^2}  \mu A_t \tan\beta\ C( m^2_{\widetilde{t}_1},
m^2_{\widetilde{t}_2}, \mu^2),
\label{botgluchar}
\end{eqnarray}
with the function $C(x,y,z)$  given by,
\begin{equation}
C(x,y,z) = \frac{\left[ (xy \ln \frac{y}{x} + yz \ln \frac{z}{y}
+ zx \ln \frac{x}{z} \right]}{(x-y)(y-z)(z-x)} >0 .
\end{equation}
This function has several simple limits:
for $x=y=z$, $C \rightarrow  1/(2x)$;  for
$x=y \gg z$, $C \rightarrow 1/x$; and, finally, for $x=y \ll z$,
$C \rightarrow \frac{1}{z} \ln \frac{z}{x}$.
Eq.~(\ref{botgluchar}) assumes that the mixing between the gaugino and
higgsino components of the chargino is small so that one chargino state
is gaugino-like with a mass $\simeq M_2$, while the other state is
higgsino-like with a mass $\mu$. We have presented these simple formulae
only to facilitate our subsequent discussion, but, as mentioned, our
computations are performed using the complete formulae from
Ref.\cite{pierce}.

\FIGURE[t]
{\hspace*{-.5cm}\epsfig{file=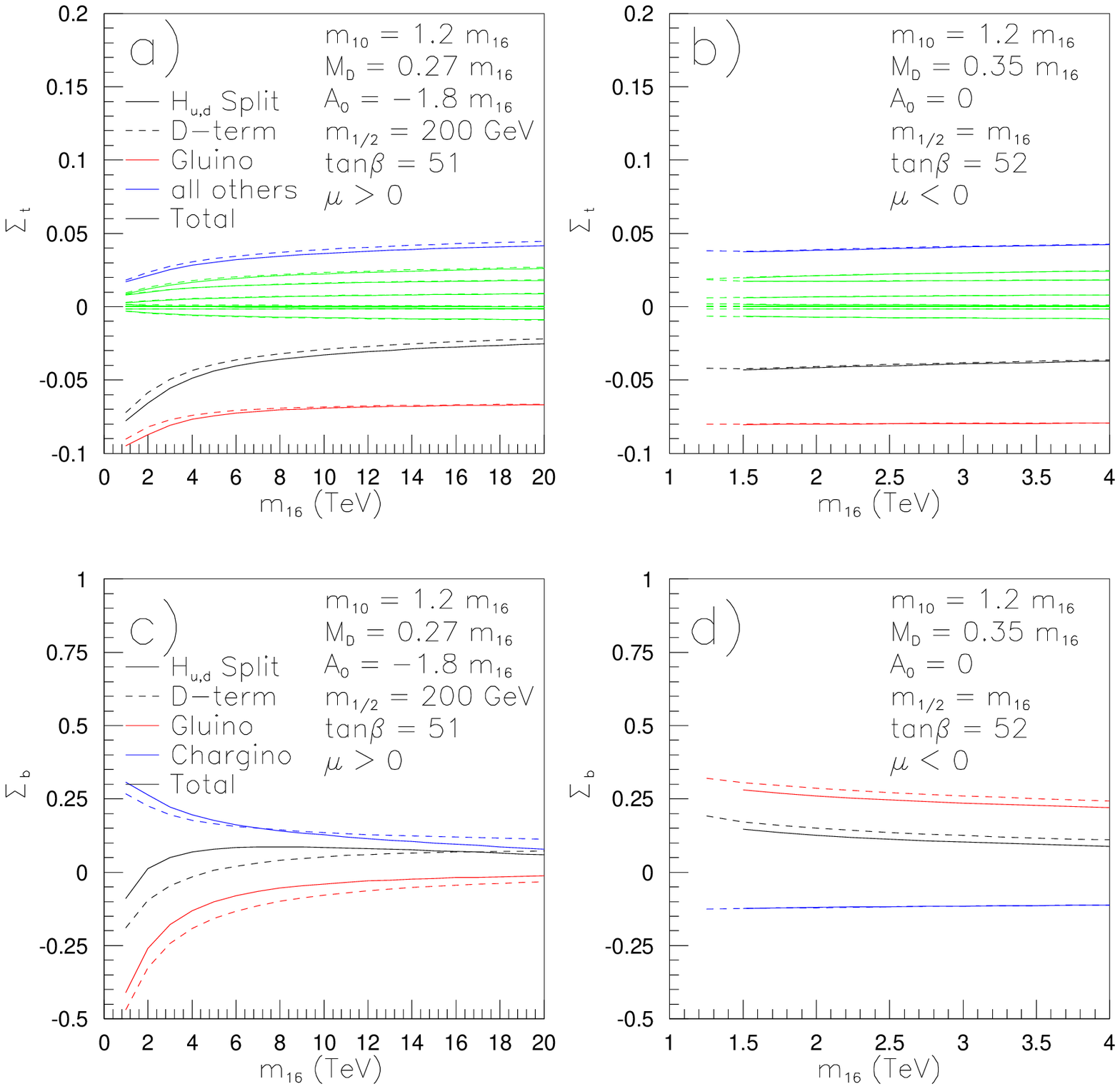,width=17cm}  
\caption{  
Various one loop contributions to the SUSY corrections to $m_t(m_t)$ and  
$m_b(M_Z)$ versus $m_{16}$, for the $DT$ and $HS$ models. The total  
corrections is shown by the dashed (solid) black line for the $DT$  
($HS$) model. various colored lines are as discussed in the text.  }  
\label{fig:loop}}  
 
For $\mu >0$ in frame {\it c}), the best Yukawa unified solutions have  
$A_t <0$. As a consequence   
the gluino and chargino loops largely cancel.   
The difference between the $DT$ and $HS$   
models is significant except for the largest values of $m_{16}$,   
and arises due to the  
differences in the values of squark masses in the two models.  
For the $\mu <0$ case in frame {\it d}),   
these contributions to $\Sigma_b$ flip sign relative to the  
$\mu >0$ case, as can be seen from  Eq.~(\ref{botgluchar}).  
This correlation with the sign of the $\mu$--term   
makes it easier to obtain Yukawa unified solutions with $\mu<0$.  
There is also only a slight   
difference in this case between the $DT$ and $HS$ model,   
reflecting our results that for $\mu <0$, there is little preference between  
the two schemes as far as Yukawa unification is concerned.  

In ISAJET, the scale $M_{GUT}$ is determined by the $Q$ value where
$g_1$ and $g_2$ have a common value. A GUT scale splitting of $g_1$ and
$g_2$ with $g_3$ is allowed since $g_3(M_{GUT})$ is determined by
evolving the central value of $\alpha_s(M_Z)$. In contrast, BDR allow an
adjustable splitting between the GUT scale gauge couplings, and find
that their best fit is obtained for a 3--4\% deviation from perfect
unification. Although this difference is amplified about four times at
$Q=M_Z$ (and so could have been a potential cause of the difference), we
see that, especially for pt. 1, in Table~\ref{tab:one} that ISAJET
results in a similar splitting between the gauge couplings. We conclude
that the additional freedom provided by $\epsilon_3$ is unlikely to
account for the difference between this analysis and that of BDR.
 
A possible criticism of our approach is that the Yukawa  
couplings are never truly unified, but only unified to within a
specified tolerance.  
To get some idea of what would be a reasonable value for this tolerance,  
we show in Fig.~\ref{fig:Rvar} the Yukawa unification  
parameter $R$ versus the inputs $m_b^{\overline{DR}}(M_Z)$ and $m_t$ for  
the $DT$ model (frames {\it a}) and {\it b})) and for the 
$HS$ model (frames {\it c}) and {\it d})) for 
several different sets of model parameters. The solid black curves 
in all four frames illustrates the variation of $R$ for the parameters 
listed in the upper frames. The dashed red and dot-dashed blue curves 
illustrate the variation of $R$ for pt.~1 of Table~\ref{tab:one} and 
pt.~5 of Table~\ref{tab:two}, respectively. For these latter two points 
parameters are such that we are close to a local minimum of $R$.  
We vary $m_b^{\overline{DR}}(M_Z)$ from 2.63 GeV to 3.03~GeV, which is 
its range quoted by the Particle Data Group, but recognize that recent 
calculations\cite{recentb} suggest that the allowed range may be just 
about half as big.  We vary the pole mass $m_t$ from 170 to 180 GeV. 
The curves are cut off if theoretical constraints that require REWSB or 
a neutralino LSP cannot be satisfied.

For the black curves, we see that as $m_t$ changes over its allowed 
range, $R$ varies by about 10\%. We have checked that in this case, 
while $f_b(M_{GUT})$ is rather stable, $f_t(M_{GUT})$ changes from 0.56 
for $m_t=175$~GeV to 0.62 for $m_t=180$~GeV, {\it i.e.} it is somewhat 
sensitive to the weak scale top Yukawa coupling, consistent with the 
finding of Ref.\cite{wells}; for $m_t$ smaller than about 175~GeV, the 
rise in the black curve in frame {\it b}) is due to a significant 
reduction in the corresponding {\it bottom} Yukawa coupling.  The 
variation in $R$ due to variation in $m_b$ is twice as much, if we allow 
$m_b^{\overline{DR}}(M_Z)$ to vary over the entire range suggested by 
the PDG compilation. For the parameters in pt.~5 of Table~\ref{tab:two}, 
the dot-dashed blue curve shows a minimum for our default choices of 
fermion mass. This should not be surprising because for this case, we 
have nearly perfect unification. The value of $R$ is rather sensitive to 
the bottom mass at the weak scale.\footnote{We have checked 
that Yukawa coupling unification occurs for any value of
$m_b^{\overline{DR}}(M_Z)$ in its allowed range. The point, however, is
that good Yukawa coupling unification appears to be possible only for
very large values of $m_{16}$. }
Finally, for pt.~1 (which is also near a local minimum of $R$) of 
Table~\ref{tab:one}, we see that $R$ changes by less than 10\% over the 
entire range of inputs for the top and bottom quark masses. Taking these 
cases to be representative, we see that while $R$ is somewhat 
sensitive to the 
weak scale fermion masses, the change in $R$ is $\alt 10\% $ 
as these are 
varied over their acceptable range, especially if we are already in the 
vicinity of a local minimum of $R$. We conclude that models with 5--10\% 
Yukawa coupling unification may be regarded as viable.\footnote{To be 
certain, we would have to scan parameters varying $m_t$ and $m_b$ over 
the allowed range and checking that perfect Yukawa unification is indeed 
obtained. Such a scan would be very time-consuming, and we have not done 
so.}

\subsection{Comparison of numerical results}  
Having compared and contrasted our procedure with the one used  
by BDR, we proceed to compare our numerical results with those  
in Ref.~\cite{bdr}. Since BDR focus on the positive $\mu$ case,   
our comparison is restricted to this.  
Both analyses agree in a number of important  
respects. The areas of agreement are:  
%
\begin{itemize}  
\item   
Yukawa unified solutions occur for $m_{16}\gg m_{1/2}$.   
\item   
The best Yukawa coupling unification occurs for boundary conditions   
close to the boundary conditions first dicussed for the RIMH scenario,   
{\it viz.} $A_0^2=2m_{10}^2=4m_{16}^2$ with  
$A_0=-2m_{16}$ fixing the sign of the $A_0$ term,  
\item   
For $\mu >0$, the $HS$ model leads to better unification of  
Yukawa couplings than  
the $DT$ model. The reason for this is discussed in the text.  
  
\end{itemize}  
  
\FIGURE[t]  
{\hspace*{-.5cm}\epsfig{file=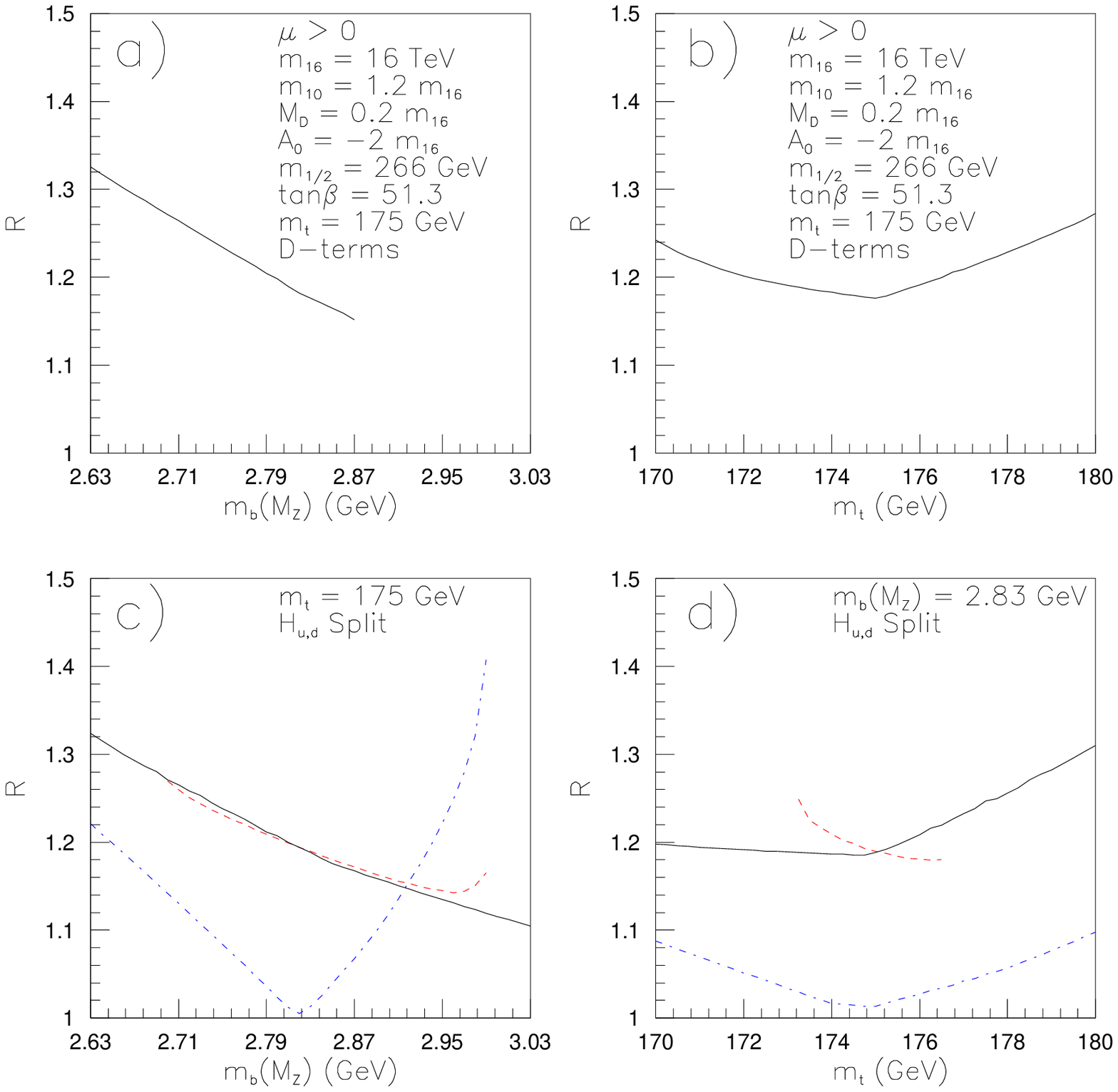,width=17cm}  
\caption{The Yukawa unification parameter $R$ versus $m_t$ (pole mass) 
and $m_b^{\overline{DR}}(M_Z)$ for different sets of parameters of 
the $DT$ and $HS$ models.  The solid black curves in all four frames 
correspond to the parameters listed in the frames {\it a}) and {\it b}) 
for the $DT$ model. The dashed red and dot-dashed blue curves in frames 
{\it c}) and {\it d}) for the $HS$ model correspond to pt.~1 of 
Table~\ref{tab:one} and pt.~5 of Table~\ref{tab:two}, respectively. The 
curves are terminated if theoretical constraints are not satisfied. } 
\label{fig:Rvar}}  
 
\noindent  
There are, however, significant differences between our results and  
those in Ref.\cite{bdr}.   
\begin{itemize}  
\item Our analysis finds that $m_{16}\sim 8-15$ TeV is needed to achieve
Yukawa coupling unification to a few per cent. 
BDR find acceptable models for
$m_{16}$ as low as $\sim 1.5-2.5$ TeV, although their
$\chi^2$ fits improve as $m_{16}$ increases.\footnote{Recently, we
have learned that BDR also find Yukawa unified solutions at
multi-TeV values of $m_{16}$, with low values of $m_{1/2}$.
These solutions give fits with $\chi^2$ as low as 0.1 to be compared to
$\chi^2$ close to 1 for the solutions
reported in Ref. \cite{bdr}. We thank R. Dermisek for
notifying us of this.} For these low $m_16$ solutions, BDR find $\chi^2$
is typically larger than about 1; since we take our weak scale
parameters to be their central experimental
values,  we do not
expect to find such solutions. 
\item BDR find Yukawa unification for $m_{1/2}\sim 200-400$ GeV,   
with $\mu \alt m_{1/2}$. These models with a higgsino-like LSP
ought to lead to a low value of neutralino relic density.
We find $m_{1/2}\sim 0-200$ GeV,  
with $\mu > m_{1/2}$. In our results, low
$|\mu|$ solutions can be generated, but the Yukawa unification never reaches
the few per cent level.
\item Our value of Yukawa couplings at $Q=M_{GUT}$ are typically  
around 0.5 to within about 10\%. This is considerably lower than   
the BDR value of the unified Yukawa coupling which ranges from   
0.63 -- 0.8 for the explicit examples in Ref.~\cite{bdr}. For RIMH-like  
boundary conditions, third generation sfermion and Higgs boson mass  
squared parameters depend exponentially on the square of the Yukawa  
couplings so that this seemingly innocuous difference in Yukawa couplings  
could have considerable impact upon the  
spectrum.   
\end{itemize}  

The low $m_{16}$ solutions found by BDR lead to rather   
different expectations for sparticle masses compared to the results
presented here. 
BDR find that  
$m_A\sim 100-200$ GeV, although values of $m_A$ up to 350 GeV are 
permitted.~\footnote{Very low values of $m_A$ may be excluded by the
upper bound on $B(B_s\to\mu^+\mu^-)$.} 
They also find $m_{\tst_1}\sim 100-300$ GeV,  
$m_{\tg}\sim 500-1000$ GeV and  
$m_{\tw_1}\sim 100-300$ GeV, where the latter particle  
has a significant higgsino component since $\mu$ is small.  
Our analysis finds $m_A\sim 1-2$ TeV, while  
$m_{\tst_1}\sim 500-2000$ GeV, $m_{\tg}\sim 300-500$ GeV  
and $m_{\tw_1}\sim 100-250$ GeV, 
where the latter is gaugino-like. In addition,  
first and second generation matter scalars are at the multi-TeV  
level, which suppresses unwanted  
$FC$ and $CP$ violating processes, but which also  
makes it difficult to obtain a reasonable dark matter relic density.  
  
   
There are also several differences in the numerical procedures   
of the two groups which could possibly account  
for the differences in the results.  
\begin{itemize}  
\item   ISAJET uses the central values of the fermion masses and gauge
couplings  while BDR make a fit to the experimentally allowed windows.
Tobe and  Wells \cite{wells} have recently suggested that Yukawa
couplings at  $Q=M_{GUT}$ may be sensitive to their values at the weak
scale; if this  is the case then, as they suggest, the degree of Yukawa
coupling  unification may indeed be affected by the fact that BDR allow
for an  experimental window for weak scale parameters. This view is
seemingly  supported by the existence of a large $\tan\beta$ infrared
fixed point for the top Yukawa  coupling, as illustrated for instance
in the second frame of Fig.~11 of  Ref.~\cite{schrempp}. However, we
note that for values of $f_t(M_{GUT})$  that we find, we are far from
this fixed point. Indeed, we have examined  Yukawa coupling unification
for the $HS$ model with $m_t=180$~GeV as  well as $m_t=172$~GeV, and
$\mu> 0$. We find that the results are  qualitatively similar to those
in Fig.~\ref{fig:HSmupdots}, except that  $R_{min}$ is attained closer
to 47--48 (51) for $m_t$ of 172~GeV  (180~GeV), and the range of $M_D$
over which the minimum is attained is  narrower for $m_t=180$~GeV.  But
the most important feature of these  scans is that we can find
solutions with better than 5\% unification for  essentially the same
range of $m_{16}$ as in Fig.~\ref{fig:HSmupdots}  for all three values
of $m_t$ that we have examined.\footnote{
We note that in all the examples of the $HS$ model in the tables  of
Ref.~\cite{bdr}, the fitted value of  $m_b(m_b)$ is larger than the
central value; this is in  qualitative agreement with the general
behaviour of $R$ in  Fig.~\ref{fig:Rvar}. } We also remark that a
larger value of $f_t$ would take  one closer to the fixed point
solution, possibly resulting in the  sensitivity suggested in
Ref.~\cite{wells}. 
We also note that for pt. 1 in 
Table~\ref{tab:one}, we found that increasing $m_t$ much beyond 
$\sim 176$~GeV lead to tachyonic $\widetilde{t}_1$, and that a change of the 
top Yukawa of this magnitude in $m_t$ resulted in a similar relative 
change in $f_t(M_{GUT})$.\footnote{The quasi-model independent
analysis  of Ref. \cite{wells} parametrizes the effects of sparticles by
adopting  different values for the threshold corrections to the
fermion  masses. Whether these values can be realized or not, depends
on the SUSY  model. For this reason, the analysis of Ref. \cite{wells}
cannot include  constraints from REWSB which are sensitive to the
details of the SUSY  framework.} We do find other points where
$f_t(M_{GUT})$ is senstive to the input value of $m_t$; however for
these cases, the unification degree is not improved by changing $m_t$
(See Fig.~\ref{fig:Rvar}, and accompanying discussion.). 
%
\item Motivated by the fact that the solutions in BDR (especially those
in Table 2 that satisfy all phenomenological constraints), we performed
a scan of the HS model for $\mu > 0$ but taking
$m_b^{\overline{DR}}(M_Z)=3.03$~GeV. As in Fig.~\ref{fig:HSmupdots}, we
find that we obtain nearly perfect unification only if $m_{16}>
8-10$~TeV. The biggest differences are that for $m_b=3.03$~GeV, $R\simeq
1$ may be obtained even if $m_{10} < m_{16}$, and the allowed range of
$M_D$ is slightly reduced. Thus, it does not appear 
to us that 
variation of $m_t$ and $m_b$ about their central values
will lead to Yukawa unified solutions with 
low $m_{16}$ as in BDR, unless we relax unification to the vicinity
of $R\sim 1.2$.
\item  
ISAJET minimizes the MSSM scalar potential at an optimal scale given by  
$\sqrt{m_{\widetilde{t}_L} m_{\widetilde{t}_R}}$ chosen to minimize  
residual scale dependent part of the one loop effective potential.  BDR  
minimize the scalar potential at $M_Z$. The difference in the  
minimization scales may considerably affect the calculation of the  
$\mu$--term (recall that BDR find the best fits for very small 
values of $|\mu|$) and the sparticle mass spectrum. 
We recall that the dominant SUSY
thresholds to $m_b$ are proportional to the $\mu$--term as can be  
seen in Eq.~(\ref{botgluchar}).   
\item  
ISAJET implements the one--loop logarithmic SUSY thresholds to   
gauge and Yukawa couplings through RGE decoupling while   
BDR implement all the susy thresholds, logarithmic and finite,   
using running parameters renormalized at the scale $M_Z$. 
\end{itemize}  
We have listed the probable   
sources of differences between the two approaches, but  
are unable to pinpoint any one  
of these as the main source of the difference from BDR  
results.  
  
\section{Summary and Concluding Remarks}  
\label{sec:conclude}  
  
We have implemented a number of improvements in the program  
ISAJET which are relevant for assessing the degree of   
third generation Yukawa coupling unification in supersymmetric  
models. Specifically,  
we examine    
$SO(10)$-inspired supersymmetric models with $D$-term  
splittings of scalar masses ($DT$ model), and with mass splittings applied  
only to the Higgs multiplets ($HS$ model).  
Using ISAJET v7.64, we have scanned a much larger  
region of parameter space than in our previous studies.   
  
Models with $\mu >0$ appear to be favored by experimental measurements  
of $BF(b\to s\gamma )$ and the muon anomalous magnetic  
moment $a_\mu$. In these models, we find that Yukawa coupling
unification is   
possible in the mSUGRA model at the 35\% level; {\it i.e.} the relative  
difference between the largest and smallest of the third generation  
Yukawa couplings at $Q=M_{GUT}$ can be as low as 35\%. In the $DT$ model,  
Yukawa unification may be possible at the 10\% level, while in the $HS$ model,   
perfect Yukawa unification is possible. However, the high degree of  
unification requires SUSY scalars to be very heavy and  
gaugino masses to be much smaller.  
The parameter space of the $HS$ model in which Yukawa  
coupling unification occurs is characterized by  
\begin{itemize}  
\item $m_{10}\sim 1.2 m_{16}$,  
\item $A_0\simeq -2 m_{16}$,  
\item $\tan\beta\sim 48-50$,  
\item $m_{16}\sim 8-20$ TeV,  
\item $m_{1/2}\sim 0-400$ GeV.  
\end{itemize}  
Parameter ranges for good unification in the $DT$ model are  
qualitatively similar.  
  
These boundary conditions  
for the soft SUSY breaking parameters are similar to those   
derived previously by Bagger {\it et al.} in the context of RIMH  
models, and was already noted by the BF and BDR analyses.  
The multi-TeV scalar masses are sufficient to yield  
a decoupling solution to the SUSY flavor and $CP$ problems.  
However, though some of the third generation scalars are considerably  
lighter, it appears that these models will require some fine-tuning  
to maintain the electroweak scale at the observed value.   
The Yukawa unified $HS$ models with $\mu >0$ yield values  
of $BF(b\to s\gamma )$ in accord with measurements, but they  
predict a value of muon anomalous magnetic moment $a_\mu$  
close to the SM value, since SUSY contributions to $a_\mu$ in this  
case essentially decouple. More problematic is the value of  
neutralino relic density $\Omega_{\tz_1}h^2$, which is usually  
much greater than 1, and thus excluded. For $m_{16}$ values  
as low as $\sim 3$ TeV, it is possible  
to obtain reasonable values of the relic density, especially  
if $m_{\tz_1}\sim m_h/2$, so that resonance annihilation  
takes place. However, unification is then possible only at  
the level of 20\%. Putting aside the issue of the relic density  
(for instance, if the LSP were unstable but long-lived),   
then for the $HS$ model with Yukawa unification,  
gluinos and charginos would be relatively light, and accessible  
to LHC experiments, and possibly even at the Tevatron or  
linear colliders. As discussed in detail in Sec.~\ref{compare}, our  
results agree in   
several respects with the recent  
work of BDR; however, there are  also significant differences, 
such as the magnitude of the $\mu$ parameter  
for which  
Yukawa coupling unification is possible. While we have not been able to  
unequivocally 
track down the precise reasons for the differences, we have attempted to  
carefully compare and contrast the two approaches to facilitate this.  
  
We have also re-examined Yukawa coupling unification for supersymmetric  
models with $\mu <0$. Both the $DT$ and the $HS$ models  
give qualitatively very similar results. Moreover,  
these results in this case are qualitatively similar  
to those derived in Refs. \cite{Baer:1999mc,Baer:2000jj}. Models with  
perfect Yukawa coupling unification can be found, but the region of  
parameter space is quite different from $\mu >0$ models. This is because  
of the sensitivity of the corrections to $m_b$ to the sign of $\mu$,  
that comes from cancellations that are necessary for Yukawa coupling  
unification if $\mu > 0$. For negative values of $\mu$, the ranges of  
parameters for which good Yukawa unification is obtained are  
characterized by,  
\begin{itemize}  
\item $\tan\beta \sim 53-56$,  
\item $m_{16}\sim m_{1/2}$,  
\item $M_D\sim (0.2-0.4)m_{16}$.  
\end{itemize}  
However, these models tend to be disfavored by $BF(b\to s\gamma )$ and  
$a_\mu$. Nonetheless, they can be in accord with the relic density  
$\Omega_{\tz_1}h^2$, since a corridor of resonance  
annihilation through $s$-channel $A$ and $H$ poles   
runs through the region where a high degree of Yukawa coupling unification  
occurs. Generally, both $m_{16}$ and $m_{1/2}$ are large  
in these models, especially for model parameter choices  
in rough accord with $BF(b\to s\gamma )$ and $a_\mu$. Thus, usually  
the SUSY particles in this case are expected to be out of reach  
of Tevatron or LC experiments, and searches at   
the LHC may also prove very challenging.

  
SUSY GUT models based on the gauge group $SO(10)$ are highly motivated.  
In many of these models, it is expected that  
third generation Yukawa  
coupling unification should occur. We have shown that supersymmetric  
models consistent with Yukawa coupling unification exist for  
both $\mu >0$ and $\mu<0$. The requirement of Yukawa coupling unification  
greatly constrains the parameter space of the models. Typically, SUSY  
scalars are very heavy, and sometimes (especially for negative $\mu$)   
all sparticles may be heavy, perhaps even beyond the reach of the LHC.  
If weak scale   
supersymmetry is found at a future collider or non-accelerator experiments,   
it will be very interesting  to see if the spectrum reflects  
qualities consistent with Yukawa coupling unification, perhaps pointing  
to $SO(10)$ as the proper gauge group for SUSY GUT models.

\section*{Acknowledgments}  
We are grateful to T.~Blazek,  
U. Chattophadyay, R. Dermisek and S.~Raby for many 
comments and suggestions.  
We especially  
thank T. Blazek for detailed discussions comparing our calculational  
algorithm with the BDR code.  We thank J.~Wells for discussion in
connection with 
Ref.\cite{wells}. 
This research was supported in part by the U.S. Department of Energy  
under contracts number DE-FG02-97ER41022 and DE-FG03-94ER40833.

%

\end{document}